\documentclass[aps,prb,twocolumn,floatfix,floats,showpacs,superscriptaddress,raggedbottom]{revtex4-1}
\usepackage{graphicx,latexsym}
\usepackage{dcolumn}
\usepackage{amsmath}

\def\mb#1{\mbox{\boldmath$#1$}}

\def\eq#1{Eq.~(\ref{#1})}
\def\fig#1{Fig.~\ref{#1}}

\usepackage{hyperref}
\hypersetup{
    pdfnewwindow=true,       
    colorlinks=true,         
    linkcolor=blue,          
    citecolor=blue,          
    filecolor=magenta,       
    urlcolor=black           
}
\pdfoutput=1
\begin{document}

\title{Time-dependent magnetotransport in an interacting double quantum wire\\
       with window coupling}

\author{Nzar Rauf Abdullah}
\affiliation{Science Institute, University of Iceland,
        Dunhaga 3, IS-107 Reykjavik, Iceland}

\author{Chi-Shung Tang}
\email{cstang@nuu.edu.tw}
 \affiliation{Department of Mechanical Engineering,
  National United University, 1, Lienda, Miaoli 36003, Taiwan}

\author{Vidar Gudmundsson}
\email{vidar@raunvis.hi.is}
 \affiliation{Science Institute, University of Iceland,
        Dunhaga 3, IS-107 Reykjavik, Iceland}

%

\begin{abstract}
We present a double quantum wire system containing a coupling
element in the middle barrier between the two parallel quantum
wires. We explicitly account for the finite length of the double
quantum wire with a time-dependent switching-on potential coupling
the double-wire system and the leads. By tuning the magnetic field
and the coupling window between the wires, we analyze the
time-dependent current and the charge distribution of the Coulomb
interacting many-electron states in order to explore inter-wire
transfer effects for developing efficient quantum interference
nanoelectronics.

\end{abstract}



\maketitle

%
%

\section{Introduction}

Quantum interference phenomena are essential when developing
mesoscale electronic devices.  Quantum confined geometries conceived
for such studies may consist of two-path
interferometers,~\cite{Schuster1997,AK2005} parallel quantum
dots,~\cite{Tang05} coupled quantum wires,~\cite{Gudmundsson06}
side-coupled quantum dots,~\cite{Orellana06,Omar2008} or Rashba
double dots in a ring.~\cite{Chen2008}  These coupled mesoscopic
systems have captured recent interest due to their potential
applications in electronic spectroscopy tools~\cite{Eugster1991} and
quantum information processing.~\cite{Schroer2006}  Nevertheless, a
study of microscopic magneotransport behavior of the transient
current flow in an interacting window-coupled double quantum wire
system is still lacking.

In the presence of a magnetic field perpendicular to the plane of
the wires, the energy spectra have been studied pointing out the
complex structure of the evanescent states of the system in
homogeneous~\cite{Barbosa97} and inhomogeneous~\cite{Korepov02}
double wires (DW). It was shown that the stepwise conductance
increasing and decreasing features can be changed by the applied
magnetic field and the height of the barrier between the
wires.~\cite{Shi97} Moreover, the dynamics of the transfer processes
for single-energy electron spectroscopy in coupled quantum states
has been considered with window coupling potential
experimentally\cite{Ramamoorthy06} and theoretically.\cite{Tang06}

In a closed time-dependently driven quantum system, the Jarzynski
relation may be derived without quantum corrections by introducing
the free-energy difference of the system between the initial and
final equilibrium state.~\cite{Mukamel2003,Monnai2005} When the
system is coupled to the reservoirs, the Jarzynski relation can be
derived using a master equation
approach.~\cite{Esposito2006,Crooks2008}  Different approaches were
proposed based on the quantum master equation (QME) to study
interaction transport
effects.~\cite{Rammer2004,Luo2007,Welack08:195315}  The time
evolution of the system described by the QME consists of two
parts: The Hamiltonian describing the system induces a unitary
evolution of the reduced density matrix, and the dissipative part
describing the properties of the environment or
reservoirs.~\cite{Lambropoulos00}

To study the time-dependent transport properties, the assumption of
Markovian dynamics and rotating wave approximation lead to different
types of master equations of the density matrix for the study of
steady-state currents by neglecting memory effects in the
system,~\cite{Kampen2001} in which the diagonal and off-diagonal
elements of the reduced density operator are
decoupled~\cite{Harbola06:235309} or assuming an infinite bias
regime.~\cite{Gurvitz1996}  However, the transient time-dependent
transport, which carries the coherence and relaxation dynamics,
cannot be generally described in the Markovian limit.  An
accurate numerical method for the nonequilibrium
time-dependent transport in the interacting nanostructures is
desirable, which can verify various approximation approaches. A
non-Markovian density-matrix formalism involving the coupled
elements should be considered based on the generalized QME
(GQME).~\cite{Braggio06:026805,Emary2007,Bednorz2008,Vidar2009,Vaz2010}
It has been confirmed that the Markovian limit not only neglects the
coherent oscillations, but also the rate at which the steady state
under this limit significantly differs from the non-Markovian
results.~\cite{Vaz2010}

In this work, we investigate how the interplay of the magnetic field
and the electron-electron ($e$-$e$) interaction affects the quantum
interference of the parallel quantum wires through a coupling window
with a time-dependent switching-on coupling to the leads.  The
central finite DW system is connected to semi-infinite leads of the
same width.  To explore the switching-on time-dependent transport
behavior through the sandwiched DW system, we shall explicitly
construct a transfer Hamiltonian that is spatially located at the
system-lead contacts and with a certain distribution in the energy
domain.  Due to the finite size of the DW system, the Coulomb
correlation could play important role in the transport.
Appropriately tuning the above physical parameters, we obtain the
transient as well as the quasi-steady state electric current using a
non-Markovian GQME method. This allows us to explore quantum
interference features of the dynamical transient currents through
the tunable window-coupled DW system.

The paper is organized as follows: In Sec.~\ref{Sec:II}, we present
the model describing the window-coupled DW system based on the GQME
theory. Section \ref{Sec:III} presents our numerical results and
physical discussion. Concluding remarks are addressed in
Sec.~\ref{Sec:IV}.

\section{Model and Theory}\label{Sec:II}

Quantum transport in an open system acted upon by a time-dependent
potential has been considered in different systems such as
time-dependent quasibound-state features,\cite{Tang1996,Tang2003}
quantum pump in Luttinger liquids,\cite{Sharma2001}
photon-associated transport in
nanostructures,\cite{Tang1999,Platero2004,Guhr2007} the Kondo effect
in a double quantum dot$-$quantum wire coupled
system,~\cite{Sasaki2006} ac-field control of spin
current,\cite{Tang2005,Amin2009} and transient current dynamics in
nanoscale junctions.\cite{Sai2007,Wang2010} The rapid progress of
nanoelectronics and information technologies has prompted intense
interest in exploiting the quantum interference transport properties
of correlated electrons, in which the coupling between the
mesoscopic subsystem could be manipulated by an applied external
magnetic field. Furthermore, the increasing interest in fast
dynamics in mesoscale systems and time-resolved detection of
electrons via a nearby detector strongly motivates investigations of
interacting time-dependent transport. It is thus warranted to
explore the magnetotransport in a central system that is
weakly-coupled to the leads by switching-on time-dependent
potentials located at the system-lead junctions.

\subsection{Single-electron Model}
One starts from an open quantum system described by a
single-electron time-dependent Hamiltonian
\begin{equation}
h(t) = h_0 + h_\mathrm{T}(t).
\end{equation}
Therein, the first term
\begin{equation}
h_0 = h_\mathrm{S} + \sum_{l=L,R} h_l
\end{equation}
indicates a disconnected single-electron Hamiltonian describing the
central system by $h_\mathrm{S}$ and the biased leads by $h_l$ with
$l$ referring to the left (L) and right (R) leads; and the second
term $h_\mathrm{T}(t)$ stands for a switching-on time-dependent
transfer Hamiltonian connecting the central system and the leads.
The $h_\mathrm{S}$ contains a disconnected Hamiltonian $h_0$ and an
envelop potential $V_{\rm DW}(\mathbf{r})$ describing the embedded
double quantum wire subsystem, namely
\begin{equation}
       h_\mathrm{S}= h_{\rm S}^0  + V_{\rm DW}(\mathbf{r}).
\label{h_S}
\end{equation}
Here $h_{\rm S}^0 = \mathbf{p}^2/2m^* + V_{\rm conf}(x,y)$ is
composed of a kinetic term with canonical momentum
$\mathbf{p}=\mb{p} + e{\bf A}$ with vector potential ${\bf A} =
(0,-By, 0)$ and a confining potential $V_{\rm conf}(x,y)=V_{\rm
c}(x) + V_{\rm c}(y)$, where $V_{\rm c}(x)$ denotes a hard-wall
confining potential at $x= \pm L_x/2$ with $L_x$ being the length of
the DW system and $V_{\rm c}(y)=\frac{1}{2}
m^*{\Omega^2_\mathrm{0}}y^2$ is a parabolic confining potential. It
is convenient to rewrite the non-perturbed single-electron central
system Hamiltonian as
\begin{equation}
 h_{\rm S}^0 = \dfrac{p_x^2}{2m^*}+ \dfrac{p_y^2}{2m^*} + \dfrac{1}{2}
 m^*\Omega_w^2 y^2 + \omega_c y p_x
\end{equation}
for defining the effective cyclotron frequency $\Omega_w^2 =
\Omega_0^2 + \omega_c^2$ in terms of the two-dimensional cyclotron
frequency $\omega_c = eB/m^*$.  The typical length scales of the
system along the $\mb{\hat{x}}$ and $\mb{\hat{y}}$ directions are
characterized by the two-dimensional magnetic length $l = (\hbar
/m^* \omega_c)^{1/2}$ and the modified magnetic length $a_w = (\hbar
/m^* \Omega_w)^{1/2}$ respectively.

Utilizing the microscopic single-electron eigenfunctions of the
system $\psi_n^\mathrm{S}(\mathbf{r})$ allows us to express the
system Hamiltonian in the spectral representation~\cite{Valeriu2009}
\begin{equation}
      h_{\rm S}=\sum_n E_n|\psi_n^\mathrm{S}\rangle\langle
      \psi_n^\mathrm{S}| ,
\label{h_Ss}
\end{equation}
where $E_n$ stands for the eigenvalues of the central system and the
dummy index $n$ refers to the quantum numbers ($n_x^\mathrm{S}$,
$n_y^\mathrm{S}$). Considering the parabolically confined
semi-infinite leads, one obtains the single-electron Hamiltonian
\begin{equation}
      h_{l}=\sum_{n_y}\int dq\, \epsilon_{n_y}^{l}(q) |\psi_{n_y,q}^l\rangle \langle\psi_{n_y,q}^l|
\label{h_l}
\end{equation}
in which $q$ stands for the continuous wave number along the
transport direction and $n_y^{l}$ denotes the transverse subband
index with $l$ referring to either of the two leads.  We assume the contact is
gradually switched on in time and calculate the time-dependent
reduced density operator of the sample using the GQME. The DW system
is coupled to the leads by introducing the off-diagonal
time-dependent transfer Hamiltonian $h_T(t) = h_T^L(t) + h_T^R(t)$,
where
\begin{equation}
      h_T^l(t)=\sum_n\int dq\: \chi^l(t) \left( T^l_{qn}|
      \psi^\mathrm{S}_n \rangle \langle \psi^l_q|+ {\rm h.c.} \right).
\end{equation}
with $T^l_{qn}$ being the coefficients connecting the eigenstates in
the system $\psi^\mathrm{S}_n$ and the leads $\psi^l_q$. Explicitly,
we express the switching-on contact function in the $l$ lead as
\begin{equation}
\chi^l(t) = \theta(t-t_0) \left[ 1 - \frac{2}{e^{\gamma (t-t_0)} +
1} \right]
\end{equation}
such that the coupling between the central DW system and the leads
is switched on at $t=t_0$ and the parameter $\gamma$ indicates the
switching rate of the coupling.  The current will flow through the
system once the switching-on contacts between the device and the
leads have been established.

\subsection{Many-electron Model}

The Coulomb interacting many-electron states (MES) of the isolated
sample are derived with the \textit{exact diagonalization}
method.~\cite{Yannouleas2007} The chemical potentials of the two
leads create a bias window which determines which MES are relevant
to the charging and discharging of the sample and to the currents,
during the transient or steady states. The many-electron Hamiltonian
\begin{equation}
 H(t) = H_0 + H_\mathrm{T}(t)
 \label{Ht}
 \end{equation}
consists of a disconnected many-electron system Hamiltonian
\begin{equation}
H_0 = H_\mathrm{S} + \sum_{l=\mathrm{L,R}} H_l
\end{equation}
and a time-dependent transfer Hamiltonian $H_\mathrm{T}(t)$. The
central system Hailtonian $H_\mathrm{S} = H_\mathrm{S}^0 +
H_\mathrm{S}^I$ contains a kinetic term $H_\mathrm{S}^0 = \sum_n E_n
d^\dagger_n d_n$ with discrete single-electron energies $E_n$ and a
Coulomb interaction term
\begin{equation}
   H_{\rm S}^I =  \sum_{n',m'}\sum_{n,m} V_{n',m'; n,m}
                                      d^\dagger_{n'} d^\dagger_{m'} d_{n}
                                      d_{m},
\end{equation}
where we have introduced the electron creation (annihilation)
operators  in the system $d^\dagger_{n}$ ($d_{n}$).  The
two-electron matrix elements
%
\begin{eqnarray}
 &&V_{n',m';n,m}  \\
 &=& \int d\mathbf{r} d\mathbf{r^{\prime}} \psi^\mathrm{S}_{n'} ({\bf r})^*
 \psi^\mathrm{S}_{m'} ({\bf r}')^* V({\bf r} - {\bf r}')
 \psi^\mathrm{S}_n ({\bf r}) \psi^\mathrm{S}_m ({\bf r}'), \nonumber
\end{eqnarray}
expressed by the single-electron state (SES) basis, are derived for
the Coulomb interaction potential
\begin{equation}
 V({\bf r} - {\bf r}') = \dfrac{e^2}{4\pi\varepsilon_0\varepsilon_r}
 \dfrac{1}{\sqrt{(x-x')^2 + (y-y')^2 + \eta ^2}}
\end{equation}
with $\varepsilon_r$ and $\eta$ being, respectively, the relative
dielectric constant of the material and the infinitesimal
convergence parameter. Below we define the dummy index
$\mb{q}=(n_y^l,q)$ and $\int d\mb{q} \equiv \sum_{n_y}\int dq$ for
simplicity. The many-electron lead Hamiltonian can be expressed in
the following form
\begin{equation}
   H_l = \int d{\mb{q}}\, \epsilon^l(\mb{q}) {c^l_{\mb{q}}}^\dagger
   c^l_{\mb{q}}.
\end{equation}
The second term in \eq{Ht} is expressed explicitly as
\begin{equation}
      H_{\rm T}^l(t)= \chi^l(t) \sum_{n}\int d{\mb{q}}\, \left[  {c^l_{\mb{q}}}^\dagger T^l_{\mb{q}n} d_n
      +   d^\dagger_n (T^l_{n\mb{q}})^* c^l_{\mb{q}}\right]
\end{equation}
describing the transfer of electrons between SES of the the system
$|n\rangle$ and the leads $|\mb{q}\rangle$ through the coupling
coefficients $T^l_{\mb{q}n}$, given by
\begin{equation}
 T^l_{\mb{q}n} =
 \int d\mathbf{r} d\mathbf{r^{\prime}} \psi^l_{\mb{q}}(\mathbf{r}')^*
 g^l_{\mb{q}n} (\mathbf{r},{\bf r'}) \psi^\mathrm{S}_n({\bf r}).
 \label{Tlqn}
\end{equation}
Therein, the coupling function
\begin{eqnarray}
      g^l_{\mb{q}n} ({\bf r},{\bf r'}) &=&
                   g_0^l\exp{\left[-\delta_x^l(x-x')^2-\delta_y^l(y-y')^2\right]}
                   \nonumber \\
                   && \times \exp{\left( -\Delta_{n}^l(\mb{q}) / \Delta
                   \right)} \label{cf}
\end{eqnarray}
containing the system-lead SES energy spread
$\Delta_{n}^l(\mb{q}) = |E_n-\epsilon^l(\mb{q})|$ making the
connection of any two SES at the contact region in the energy
domain.~\cite{Vidar2009} The spatial coupling range in the leads is
governed by $\delta_x^l$ and $\delta_y^l$. We have considered the
energy interval $[\mu_R-\Delta, \mu_L + \Delta]$ to define an active
window in the energy domain $\Delta_E = \Delta \mu +2 \Delta$ that
involves all the possible states in the central system that are
relevant to the transport. It should be mentioned that only the
transverse part of the wave function in the semi-infinite leads is
normalizable. To get rid of all length scales variation with
magnetic field, one needs to fix $g_0^l a_w^{3/2}$ in units of
energy and then calculate $g_0^l$.

\subsection{GQME Formalism}

In this subsection, we formulate the time evolution of the MES when
the system contains a number of electrons for the study of
interacting time-dependent transport properties based on the GQME
formalism.\cite{Breuer2002} To take into account the many electrons
in the system, we construct a Fock space by selecting the number of
the $N_{\mathrm{SES}}$ lowest single-electron states and the
$N_{\mathrm{MES}}=2^{N_{\mathrm{SES}}}$ many-electron states within
the active window $\Delta_E$. In the occupation representation
basis, the noninteracting MES
\begin{equation}
      |\alpha\rangle =
      |i^\alpha_1,i^\alpha_2,\dots,i^\alpha_n,\dots, i^\alpha_{N_{\rm SES}}\rangle
\end{equation}
contains the labels $i^\alpha_n=0,1$ indicating the occupation of
the $n$-th SES of the isolated central system within the active
window. The corresponding energy of the noninteracting MES ${\cal
E}_{\alpha}=\sum_n E_ni_n^{\alpha}$ can be obtained by summing over
the occupied SES.

The time-evolution of the many electron system under investigation obeys the Liouville-von
Neumann (quantum Liouville) equation~\cite{Esposito2009}
\begin{equation}
 \frac{dW(t)}{dt}= -\frac{i}{\hbar}\left[H(t),W(t)\right],
\end{equation}
where the full density operator $W(t)$ can be operated upon by a
projector to yield the reduced density operator (RDO) by taking
trace over the Fock space in the leads $\rho(t)={\rm Tr}_\mathrm{L}
{\rm Tr}_\mathrm{R} W(t)$,  with $\rho(t_0) =
\rho_\mathrm{S}$.~\cite{Haake1971} The initial condition $W(t<t_0)$
= $\rho_\mathrm{L}\rho_\mathrm{R}\rho_\mathrm{S}$ is in terms of the
equilibrium RDO of the disconnected lead $l$ with chemical potential
$\mu_l$, given by
\begin{equation}
      \rho_l=\frac{e^{-\beta (H_l-\mu_l N_l)}}{{\rm Tr}_l \{e^{-\beta(H_l-\mu_l N_l)}\}}
\end{equation}
with $l$ referring to the $L$ and the $R$ leads. This allows us to find
the equation of motion for the RDO of the following
form\cite{Haake1973}
\begin{equation}
\frac{d{\rho}(t)}{dt} = -i{\cal L}_{\rm eff}\rho(t) +\int_{t_0}^{t}
dt' {\cal K}(t,t')\rho (t'),
\end{equation}
where ${\cal L}_{\rm eff}$ stands for the effective Liouvillian and
${\cal K}(t,t')$ denotes the integration kernel.~\cite{Haake1973}

Using the \textit{exact diagonalization} method,  we diagonalize the
interacting system Hamiltonian $H_{\rm S}$ in the MES basis of the
noninteracting system $\{|\alpha\rangle\}$ in the Fock space. Since
we are dealing with an open system with variable electron number,
one has to include all sectors containing zero to $N_{\mathrm{SES}}$
electrons. This yields a new interacting MES basis $\{|\mu)\}$ with
\begin{equation}
 |\mu) = \sum_{\alpha} {\cal U}_{\mu\alpha}|\alpha\rangle
\end{equation}
connected by the $N_{\mathrm{MES}}\times N_{\mathrm{MES}}$ unitary
transformation matrix ${\cal U}_{\mu\alpha}$. A basis transformation
of the interacting many-electron coupling matrix $\widetilde{{\cal
T}}^l(\mb{q}) = {\cal U}^\dagger{{\cal T}}^l(\mb{q}){\cal U}$  and
the insertion of the diagonalized matrix representation of the
interacting $H_\mathrm{S}$ allows us to obtain the RDO in the
interacting MES basis $\widetilde{\rho} = {\cal U}^\dagger\rho{\cal
U}$.  Expressing the interacting many-electron coupling matrix
$\widetilde{\cal T}$ in the interacting MES
\begin{equation}
      \widetilde{\cal T}^l(\mb{q})
      =\sum_{\mu,\nu}\widetilde{\cal T}_{\mu\nu}^l(\mb{q})
      |{\bf \nu})({\bf \mu}|
\label{Toperator}
\end{equation}
with $\widetilde{\cal T}_{\mu\nu}^l(\mb{q}) = \sum_n
T^l_{n\mb{q}}({\mu} |d_n^{\dagger}|{\nu})$ in terms of
 the
single-electron coupling matrix $T^l_{nq}$, one can obtain the
transformed GQME
\begin{eqnarray}
 \frac{d{\widetilde{\rho}}(t)}{dt}
 &=& -\frac{i}{\hbar}\left[ H_{\rm S},\widetilde{\rho}(t)\right]\\
 && -\frac{1}{\hbar^2}\sum_{l=L,R}\chi^l(t) \int d\mb{q}\: \left( \left[\widetilde{\cal
T}^l(\mb{q}),\Omega_{\mb{q}}^l(t)\right] + {\rm h.c.}\right) .
\nonumber \label{TGQME}
\end{eqnarray}
Here we have defined the effective interacting coupling operator
\begin{eqnarray}
 \Omega_{\mb{q}}^l(t) &=& U_\mathrm{S}^\dagger (t)
 \int_{t_0}^tds\:\chi^l(s)\Pi_{\mb{q}}^l(s) \nonumber \\
  &&\times \exp{\left[ -\frac{i}{\hbar}(t-s) \epsilon^l(\mb{q})  \right]} U_\mathrm{S}(t)
\end{eqnarray}
with
\begin{eqnarray*}
 \Pi_{\mb{q}}^l(s) &=& U_\mathrm{S}(s)
      \left[ \left(\widetilde{\cal T}^l\right)^{\dagger} \widetilde{\rho}(s)\left[ 1 -
      f^l\left( \epsilon(\mb{q})\right)
      \right] \right.  \\
      && - \left. \widetilde{\rho}(s)\left(\widetilde{\cal T}^l\right)^{\dagger}
      f^l\left( \epsilon(\mb{q}) \right)
      \right]
      U_\mathrm{S}^\dagger(s),
\end{eqnarray*}
in which $U_\mathrm{S}(t)=e^{iH_\mathrm{S}(t-t_0) / \hbar}$ denotes
the time evolution operator and $f^l\left( \epsilon(\mb{q}) \right)
= \{\exp[\epsilon(\mb{q})-\mu_l]+1\}^{-1}$ indicating the Fermi
function in the $l$ lead at $t=t_0$.  In the numerical calculation
we shall select $t_0=0$ for convenience.

Taking the statistical average over the Fock space $\langle
\hat{Q}_{\rm S}(t)\rangle = {\rm Tr} \{ W(t) \hat{Q}_{\rm S} \}$ of
the charge operator $\hat{Q}_{\rm S} = e \sum_n d_n^{\dagger} d_n$
in the coupled central system and using the identity
$\widetilde{\rho}(t)={\rm Tr}_{\rm L}{\rm Tr}_{\rm R}\{W(t)\}$, one
may express the statistical averaged time-dependent charge as
\begin{equation}
 \langle \hat{Q}_{\rm S}(t)\rangle = e \sum_n \sum_{\mu} i^{\mu}_n \,
 \left( \mu \left| \widetilde{\rho}(t) \right| \mu \right) .
\end{equation}
This allows us to define the time-dependent net charge current
flowing through the central DW system
\begin{equation}
 I_Q(t) = \frac{d\langle \hat{Q}_\mathrm{S}(t) \rangle}{dt}  = I_L(t) - I_R(t).
 \end{equation}
The charge current injected from the $l$ lead to the system is given
by
\begin{equation}
    I_l(t) = e\sum_n \sum_{\mu} i^{\mu}_n \,
      \frac{d{\widetilde{\rho}}^{\:l}_{\mu\mu}}{dt} ,
\label{Il}
\end{equation}
in which we express the current in terms of the time derivative of
the reduced density matrix elements in the interacting MES basis:
\begin{equation*}
      \frac{d{\widetilde{\rho}}^{\:l}_{\mu\mu}}{dt} =
      -\frac{\chi^l(t)}{\hbar^2}
      \int d\mb{q}\: \left( \mu \left| \left[\widetilde{\cal
T}^l(\mb{q}),\Omega_{\mb{q}}^l(t)\right] + {\rm h.c.}\right| \mu
\right) . \label{current}
\end{equation*}
It is straight forward to obtain the interacting many-electron
charge distribution in the DW system
\begin{equation}
  Q({\bf r},t) = e\sum_{n',n} \psi^*_{n'}({\bf r}) \psi_n({\bf r})
                  \sum_{\mu,\nu} \widetilde{\rho}_{\mu\nu}(t) (\mu |d^\dagger_{n'} d_n|\nu).
      \label{Qxy}
\end{equation}
Below we shall show our numerical results of the net time-dependent
charge current $I_Q(t)$ through the central DW system. It is an
algebraic sum of the left current $I_L(t)$ (indicating the charge
current from the left lead to the right lead) and the right current
$I_R(t)$ (indicating the charge current from the system to the right
lead). We shed light on the transport dynamics by analyzing the
time-dependent many-electron charge distribution $Q({\bf r},t)$ in
real space.

\section{Results and Discussion}\label{Sec:III}

We numerically solved the GQME to investigate the dynamical
time-dependent magnetotransport of electrons through a central
finite system of length $L_x=300$~nm with magnetic length
$l = (h/(eB))^{1/2} = 25.67
[B({\rm T})]^{-1/2}$~nm. The central system is transversely confined
by a parabolic potential with characteristic energy $\hbar\Omega_0 =
1.0$~meV. This supplies the modified magnetic length
\begin{align}
a_w &= \left(\frac{\hbar}{m^*\Omega_0}\right)^{1/2}
    \left(\frac{1}{1+(eB/(m^*\Omega_0))^2}\right)^{1/4}\nonumber\\
    &= \frac{33.74}{ \sqrt[4]{1+2.982[B({\rm T})]^2} }\ {\rm nm},
\end{align}
and the typical width of the confined system for the lowest subband
electron is $L_y \approx 67.5$~nm. We assume GaAs parameters with
electron effective mass $m^*=0.067m_e$ and the background relative
dielectric constant $\varepsilon_r = 12.9$.

\begin{figure}[htbq]
      \includegraphics[width=0.42\textwidth,angle=0]{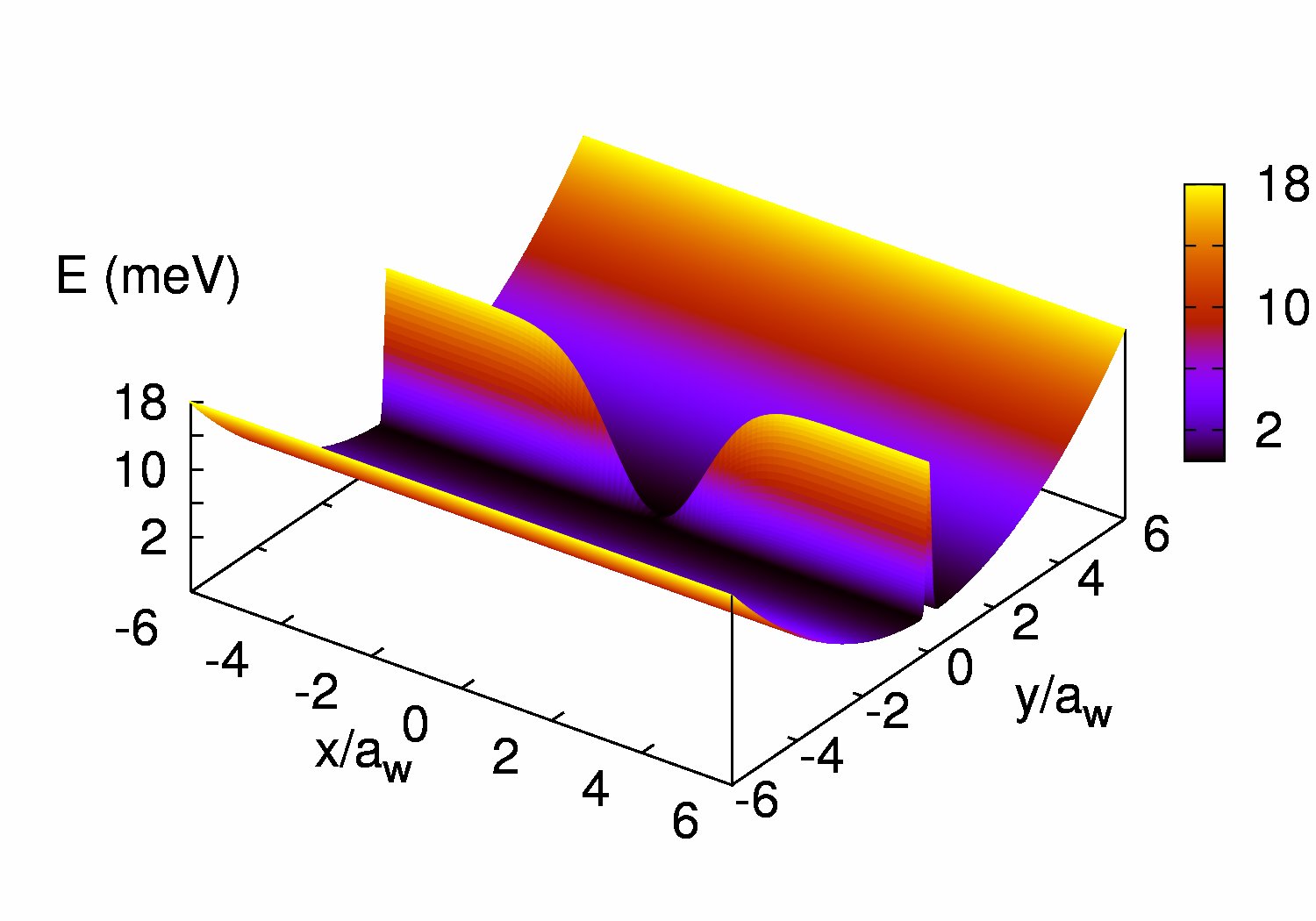}
      \caption{(Color online) Schematic illustration of the potential defining the
      window-coupled DW system. $\hbar\Omega_0 = 1.0$~meV, $B=0$~T, $a_w=33.74$~nm.}
      \label{System}
\end{figure}
Figure \ref{System} schematically illustrates the window-coupled DW
system scaled by $a_w$.  The embedded DW system is described by
$V_{\rm DW}(\mathbf{r}) = V_{\rm MB}(y) + V_{\rm CW}(x,y)$ that
contains a middle barrier
\begin{equation}
V_{\rm MB}(y) = V_0\; {\rm exp}{(-\beta_0^2y^2)}
\end{equation}
with $V_0=18.0$~meV and $\beta_{0}=0.3$ nm$^{-1}$, as well as a
coupling window potential
\begin{equation}
V_{\rm CW}(x,y) = -V_0\; {\rm exp}{(-\beta_x^2 x^2-\beta_y^2 y^2)}.
\end{equation}
The coupling constant $g_0^l a_w^{3/2}= 60\ {\rm meV}$, and the
contact size parameter $\delta_x^l =\delta_y^l =4.4\times
10^{-4}$~nm$^{-2}$

In the following calculations, the temperature of the reservoirs is
fixed at $T=0.5$~K, and the states within the bias window before
switching-on the coupling are assumed to be unoccupied. The coupling
between the DW system and the leads is characterized by the
switching rate $\gamma = 1.0$~ps$^{-1}$, and the nonlocal coupling
strength is fixed as $\Gamma^l = 4 g_0^l a_w^{3/2} /
(\delta_x^l\delta_y^l)^{1/2} = 54.5$~meV$\cdot$nm$^2$. The bias
voltage is fixed leading to a bias window $eV_{\rm bias} = \Delta\mu
= 0.9$~meV, and the extension parameter $\Delta=0.3$~meV is selected
referring to a window of relevant states $\Delta_E = \Delta\mu +
2\Delta = 1.5$~meV.

\begin{figure}[htbq]
 \includegraphics[width=0.23\textwidth]{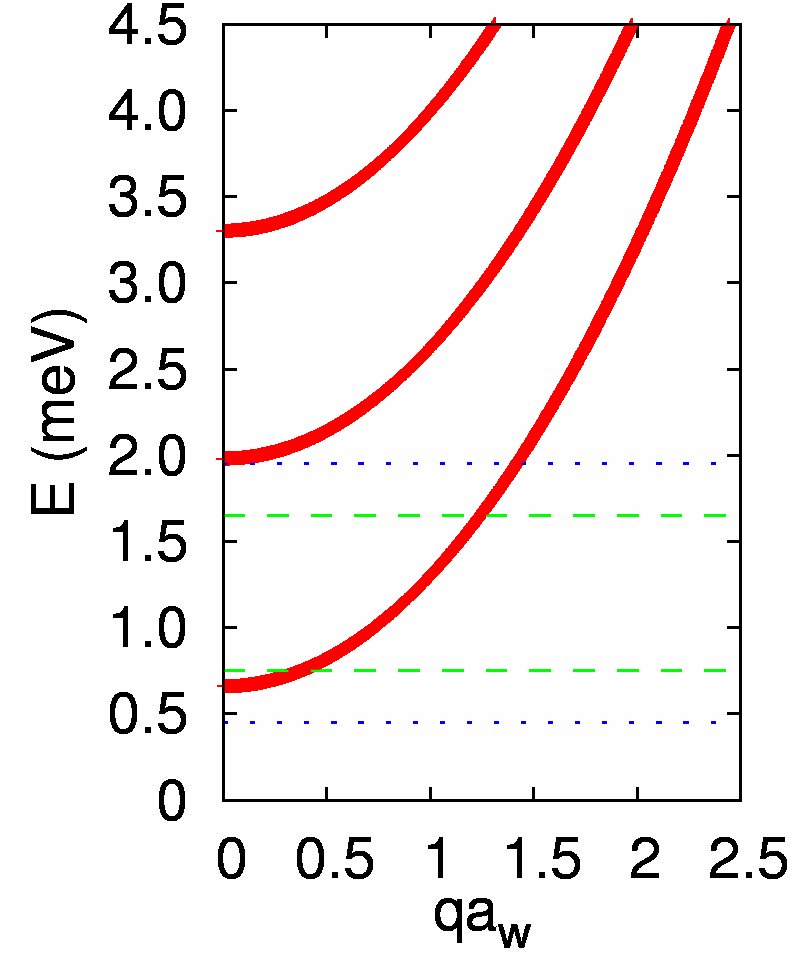}
  \includegraphics[width=0.23\textwidth]{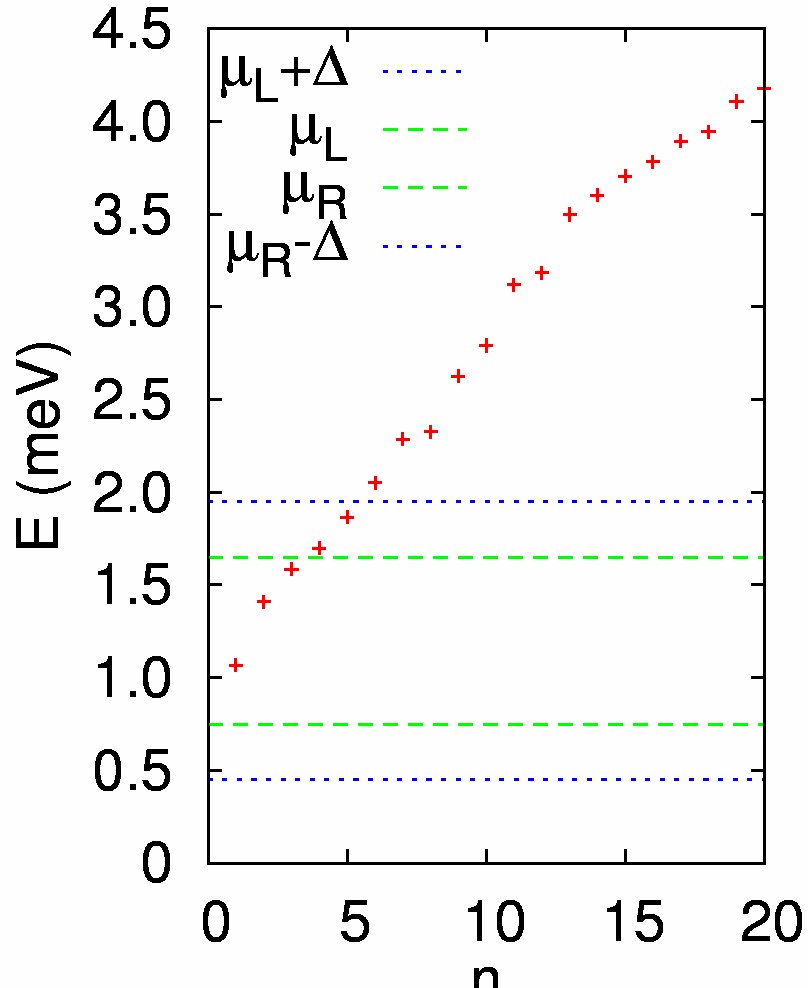}
\caption{(Color online) Energy spectrum of the leads (solid red)
versus wave number $q$ (left panel); and energy spectrum of the
window-coupled DW system (cross dot) versus the SES number $n$
(right panel). Magnetic field $B=0.5$~T, and the chemical potentials
in the the leads are $\mu_L=1.65\ {\rm meV}$ and $\mu_R=0.75\ {\rm
meV}$ (dashed green) such that $\Delta\mu = 0.9$~meV. The window of
relevant states $\Delta_E = 1.5$~meV is defined by the dotted blue
lines.} \label{ES-S}
\end{figure}

The energy spectrum of the leads as a function of wave number $q$
scaled by $a_w^{-1}$ is shown in the left panel of \fig{ES-S}. The
bias window $\Delta \mu$ is located in the first subband, whereas
the extended active bias window covers the evanescent modes below
the first subband and the threshold of the second subband. The
energy spectrum of the window-coupled DW system as a function of the
single electron number $n$ is shown in the right panel of \fig{ES-S}
containing five SESs in the window of relevant states $\Delta_E$;
the three lowest states are in the bias window $\Delta\mu$ whereas
the two highest states are in the upper extended window $[\mu_L,
\mu_L+\Delta]$.

\begin{figure}[htbq]
 \includegraphics[width=0.45\textwidth]{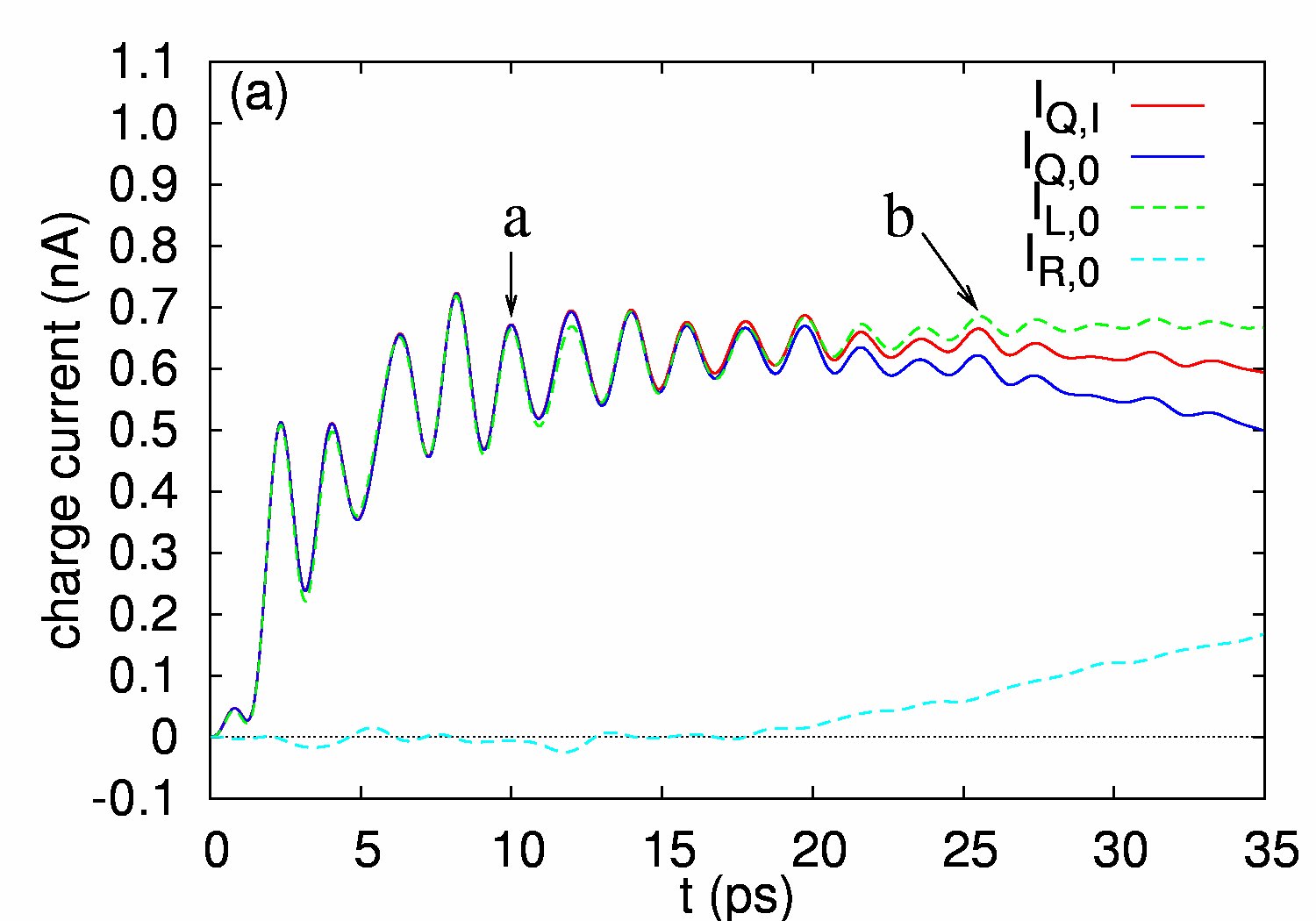}
  \includegraphics[width=0.45\textwidth]{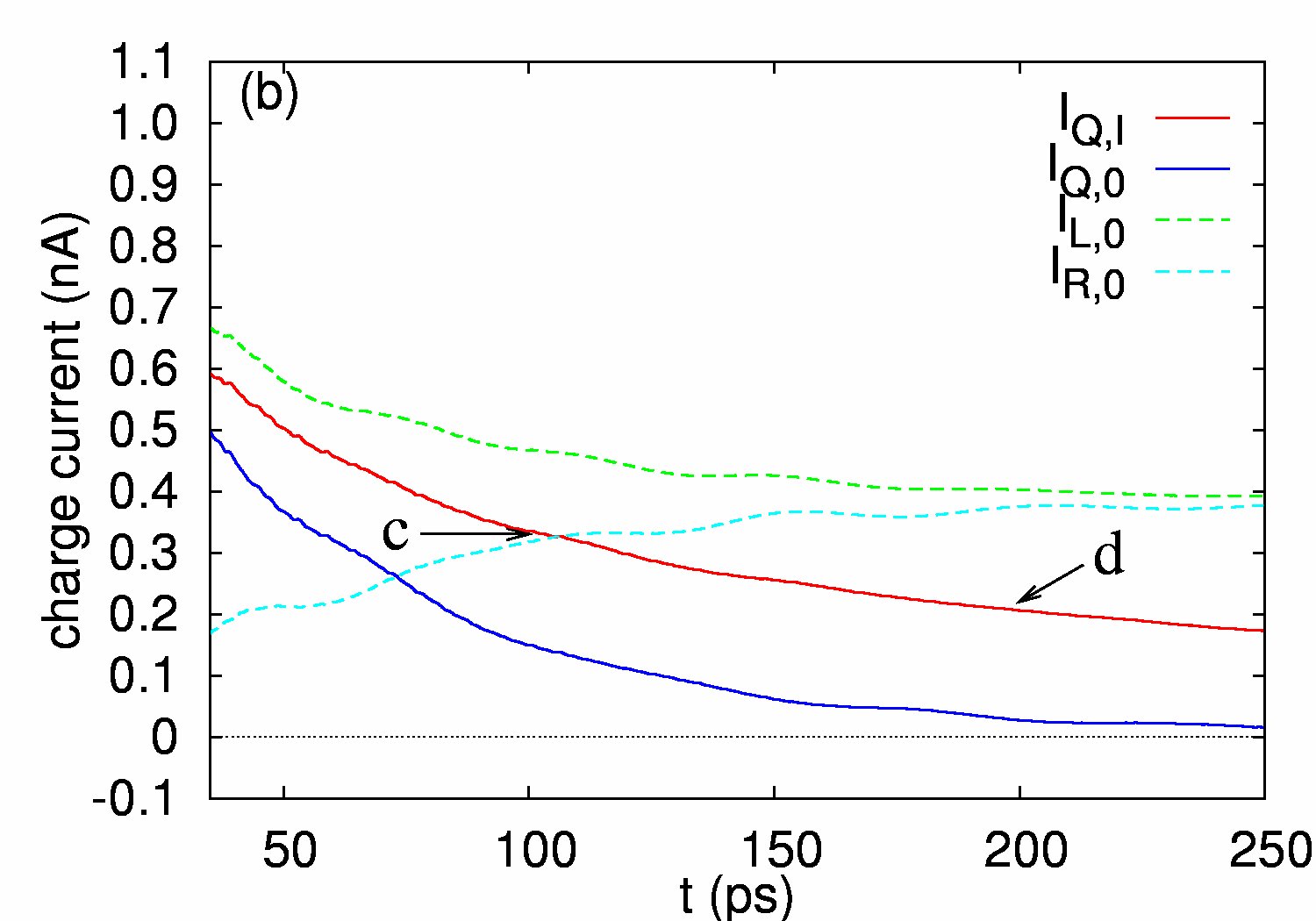}
\caption{(Color online) The interacting net current $I_{Q,I}$ (solid
red), the noninteracting net current $I_{Q,0}$ (solid blue), the
noninteracting left current $I_{L,0}$ (dashed green), and the
noninteracting right current $I_{R,0}$ (dashed light-blue) are
plotted as a function of time: (a) short-time response; (b)
long-time response. $B = 0.5$~T, $L_{\rm w} = 100$~nm, the coupling
constant $g_0^l a_w^{3/2}= 60\ {\rm meV}$, and the contact size
parameters $\delta_x^l =\delta_y^l =4.4\times 10^{-4}$~nm$^{-2}$.}
\label{I-B05}
\end{figure}

In \fig{I-B05}, we show the time-dependent charge current for the
case of magnetic field $B=0.5$~T with and without $e$-$e$
interaction, denoted by $I_{Q,I}$ and $I_{Q,0}$ respectively. The
noninteracting left and the right currents are also presented for
comparison, denoted by $I_{L,0}$ and $I_{R,0}$ respectively.  We
have selected $\beta_{x}=0.02$ nm$^{-1}$ and $\beta_{y}=\beta_{0}$
such that the length of the coupling window $L_{\rm w}$ is $100$~nm.
In addition, the coupling constant is $g_0^l a_w^{3/2}= 60\ {\rm
meV}$ and the contact size parameters are $\delta_x^l =\delta_y^l
=4.4\times 10^{-4}$~nm$^{-2}$ such that the coupling strength
$\Gamma^l = 54.5$~meV$\cdot$nm$^2$ and the effective lengths of the
system-lead coupling potential are $L_{c,x}^l=L_{c,y}^l\approx
95$~nm. Below we shall show that the time-dependent charge current
manifests different transport mechanisms in the short-time and the long-time
response.

In the short-time response regime, shown in \fig{I-B05}(a), the
time-dependent charge current is increased and manifests rapid
oscillation with period $\tau_s\approx 1.9$~ps exhibiting quantum
interference dominant features.  In this regime, the noninteracting
approach could be a good approximation for analyzing the transient
time-dependent transport properties. In this short-time regime, the
interacting and the noninteracting currents are almost the same
before $20$~ps with negligible right charge current implying
effective charging and quantum interferance dominant transport
feature. The right charge current is significantly increased after
$20$~ps. At around $t=35$~ps, the difference between the interacting
and noninteracting currents becomes $0.1$~nA (the Coulomb correction
is $\sim 10\%$), and the right charge current is increased to
$0.18$~nA.

In the long-time response shown in \fig{I-B05}(b), the charge
current displays slow quasi-periodic oscillation with period
$\tau_l\approx 39$~ps approaching a steady current. The
slow oscillation behavior in the time-dependent current implies that
the quantum interference feature is suppressed whereas the Coulomb
interaction effect is enhanced.  At time $t=250$~ps, the interacting
steady current ($\sim 0.17$~nA) is much higher than the
noninteracting steady current ($\sim 0.015$~nA).   The mean charge
of the DW system is monotonically increased in time (not
shown),\cite{Vidar2010} and the mean charge of the interacting MES
($\sim 0.8e$) is approximately twice that of the steady mean charge of
the noninteracting MES ($\sim 0.4e$).  This indicates that the
empty-state initial condition ensures that the Coulomb
interaction facilitates to drag the electron dwelling in the DW
system through the window of relevant states, and thus enhances the steady
current.

\begin{figure}[htbq]
  \includegraphics[width=0.22\textwidth]{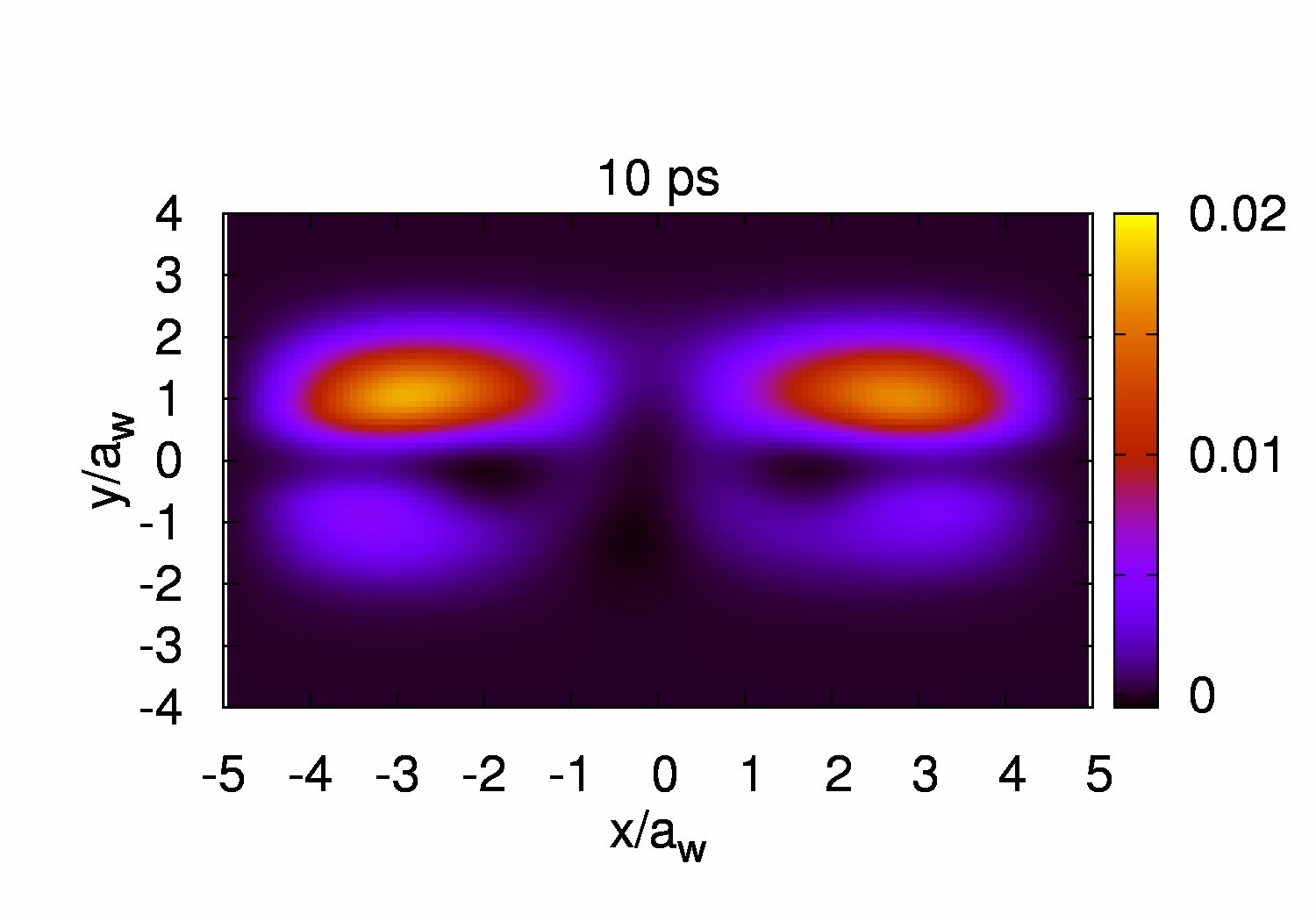}
  \includegraphics[width=0.22\textwidth]{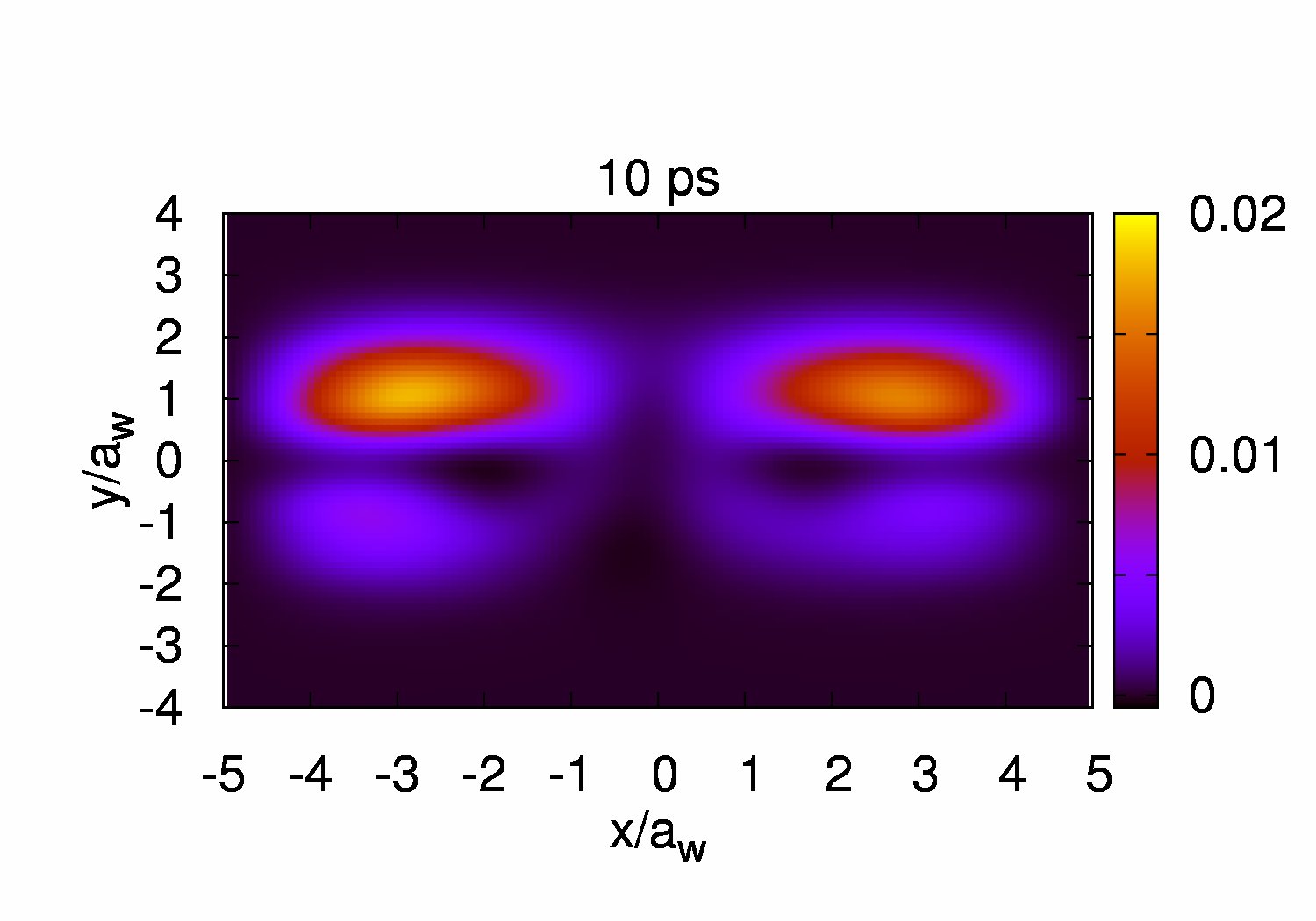}\\
  \includegraphics[width=0.22\textwidth]{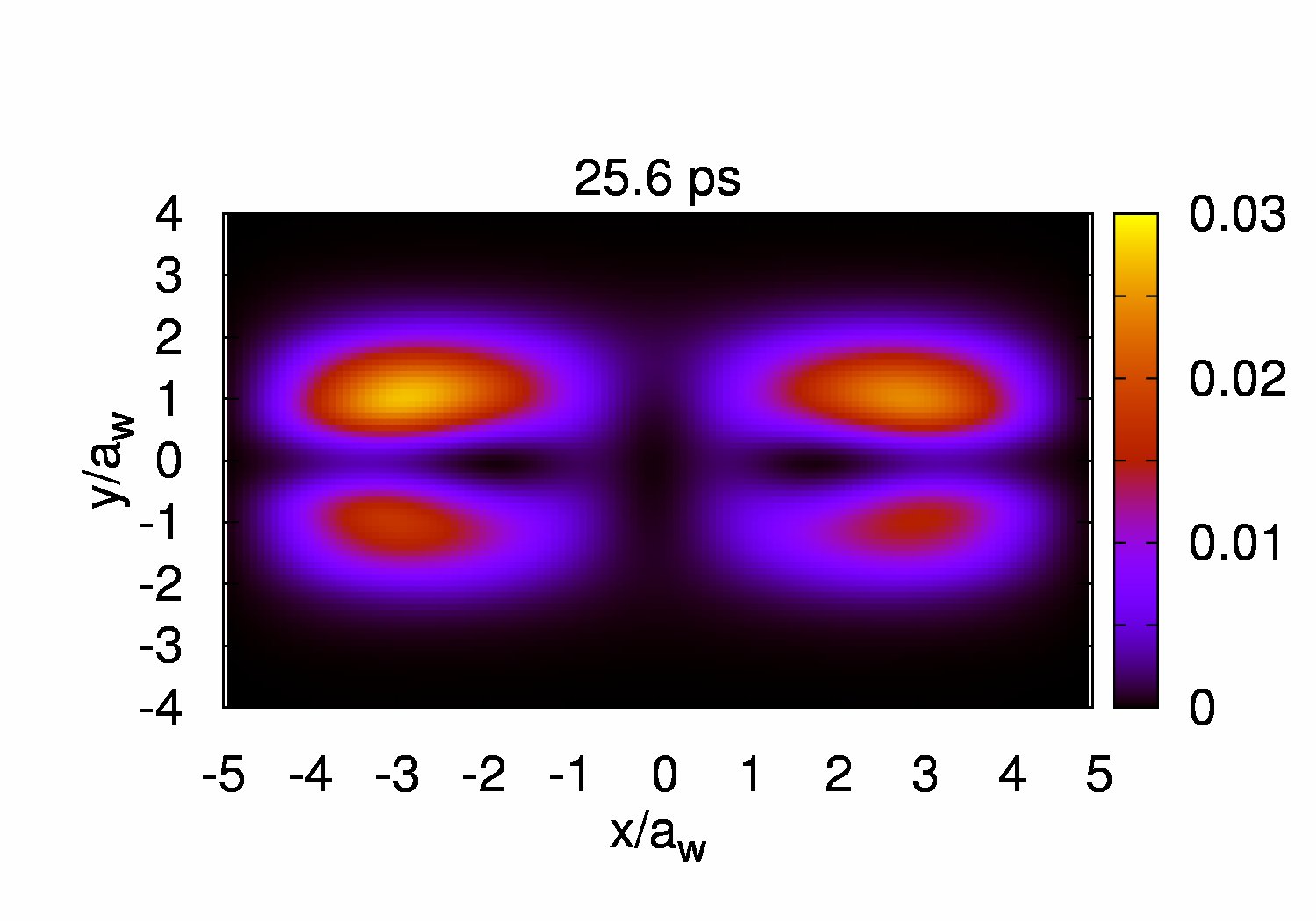}
  \includegraphics[width=0.22\textwidth]{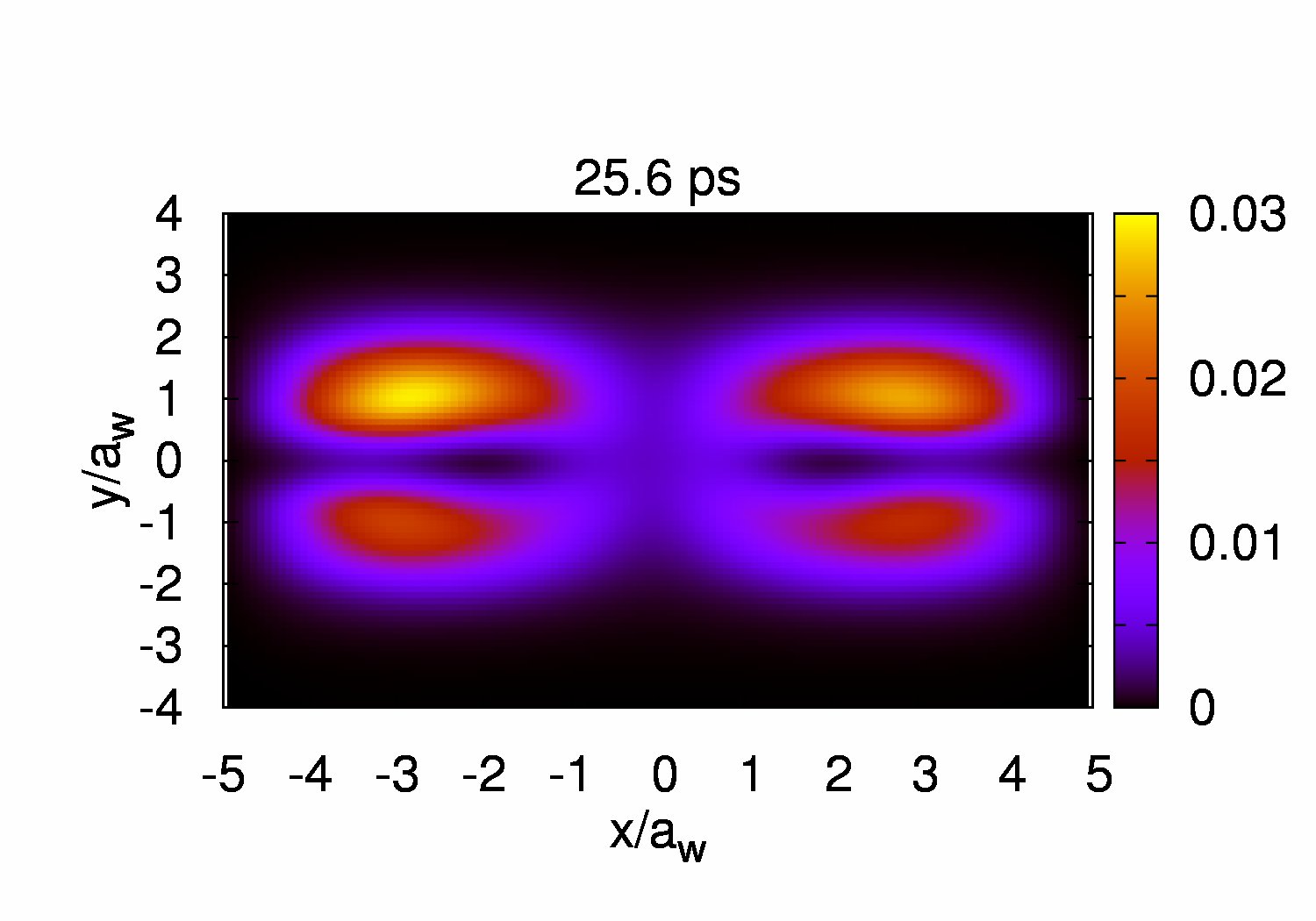}\\
  \includegraphics[width=0.22\textwidth]{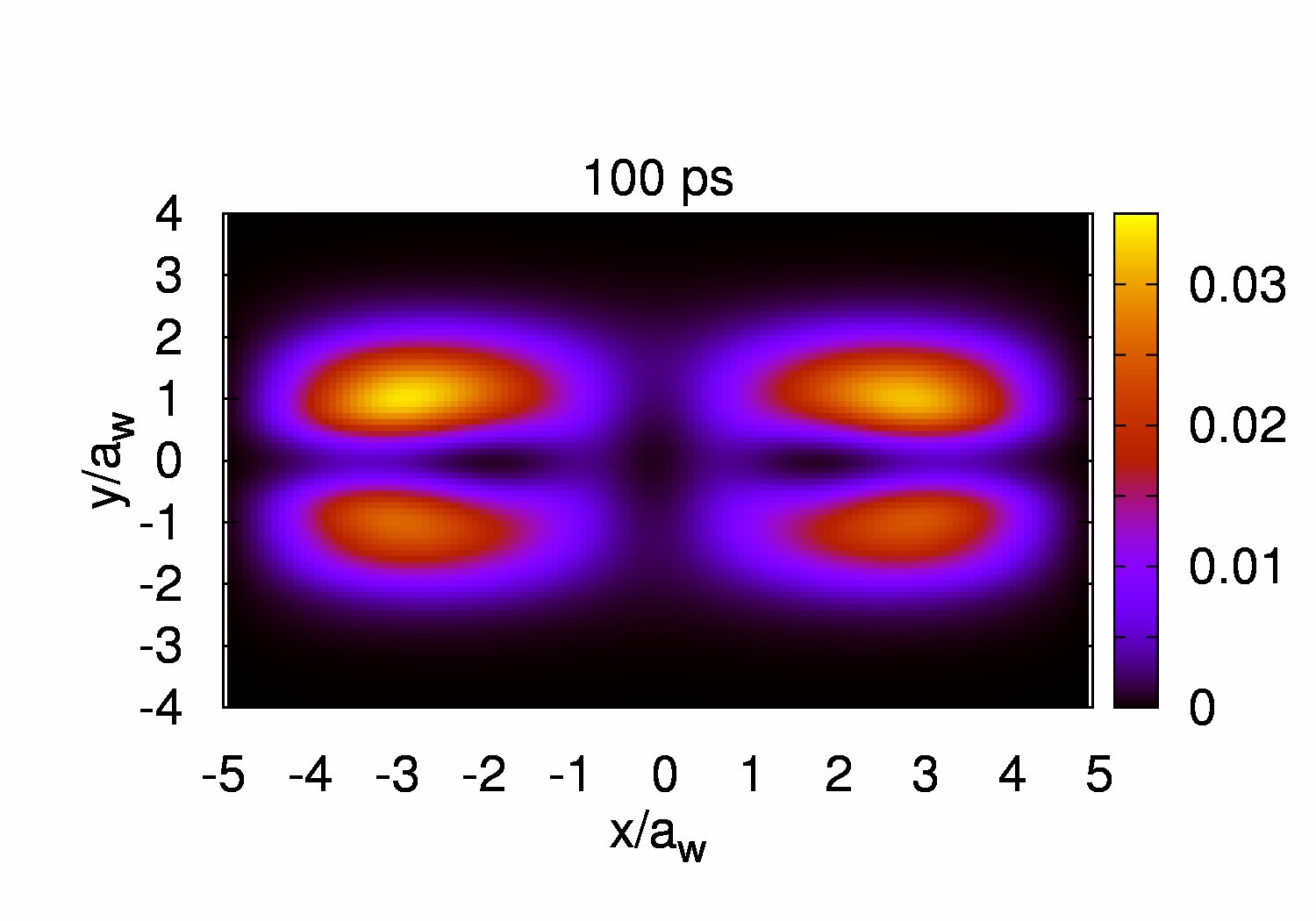}
  \includegraphics[width=0.22\textwidth]{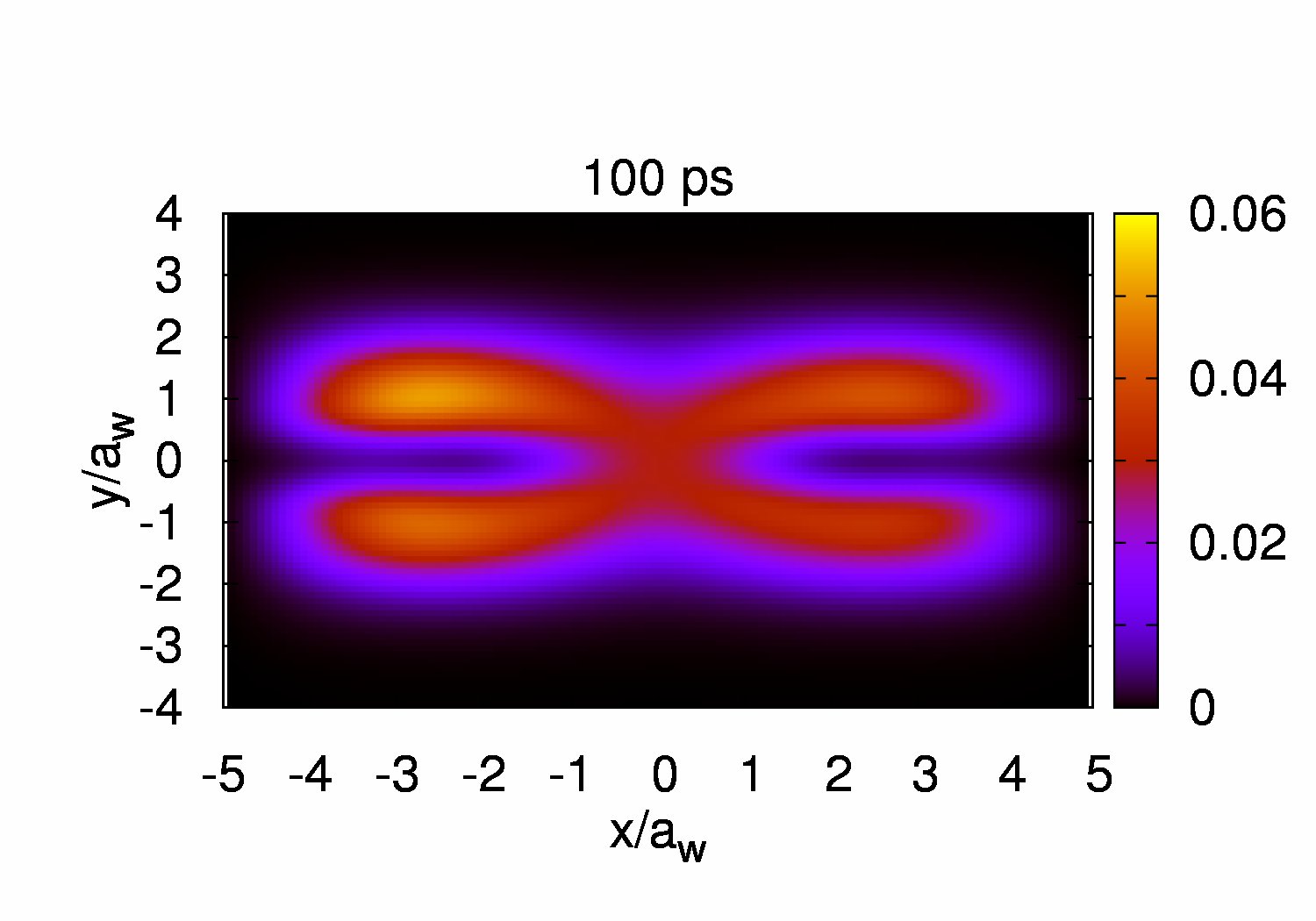}\\
  \includegraphics[width=0.22\textwidth]{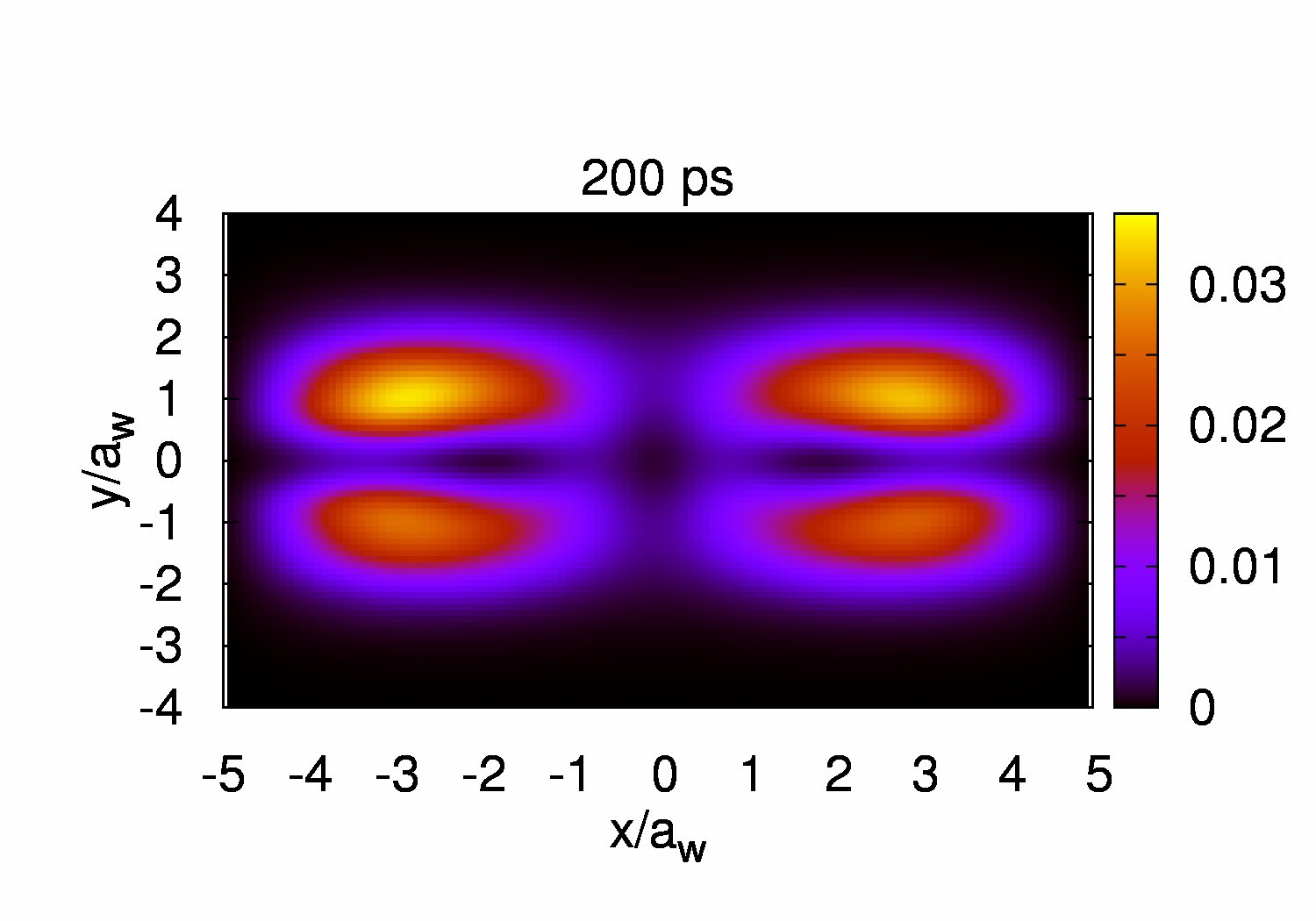}
  \includegraphics[width=0.22\textwidth]{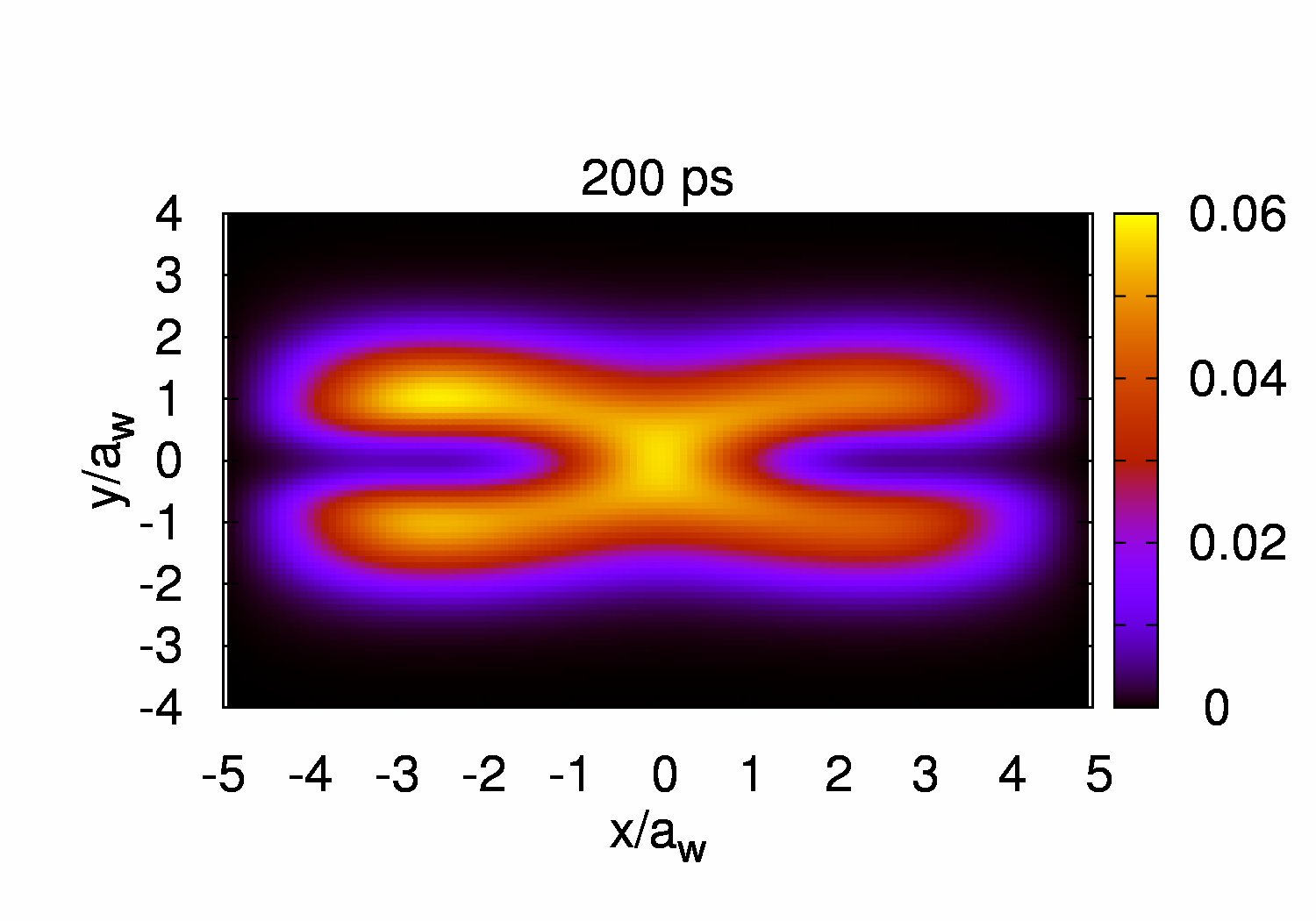}
\caption{(Color online) The many-electron charge density for the
noninteracting (left panel) and interacting system (right panel) for
$B=0.5$~T. The other parameters are the same as \fig{I-B05}.}
\label{QB0.5}
\end{figure}
In order to get better understanding on the transient dynamical
transport, we present the spatial distribution of the many-electron
charge at $t=10$, $25.6$, $100$, and $200$~ps in \fig{QB0.5},
labeled by $\bf{a}$-$\bf{d}$ in \fig{I-B05}, respectively. When the
system-lead coupling is switched-on with forward bias, the electrons
are incident from the left lead into the system with
transversely symmetric distribution (not shown). At around $t\simeq 10$~ps, the
electrons located in the lower wire favorite to make inter-wire
backward scattering to the upper wire, exhibiting a fully quantum
mechanical feature. Later on, the electrons perform an
opposite inter-wire backward scattering feature to the lower wire at
$t \simeq 25$~ps, and this feature is only slightly enhanced by the
Coulomb interaction. However, in the long-time response regime, the
inter-wire scattering forward and backward effects are both
enhanced.  At around $t= 100-200$~ps, the noninteracting
window-coupled DW forms a quasi-isolated four cavities, the window
coupling effect is significantly enhanced by the Coulomb
interaction.  It is interesting that the electron can form a
quasibound state in the coupling window at $t\simeq 200$~ps.  When
the DW system approaches steady-state transport in the long-time
response regime, the total charge in the system is $0.4e$ for
noninteracting and $0.8e$ for interacting DW system exhibiting
significant charge accumulation behavior.

\begin{figure}[htbq]
 \includegraphics[width=0.45\textwidth]{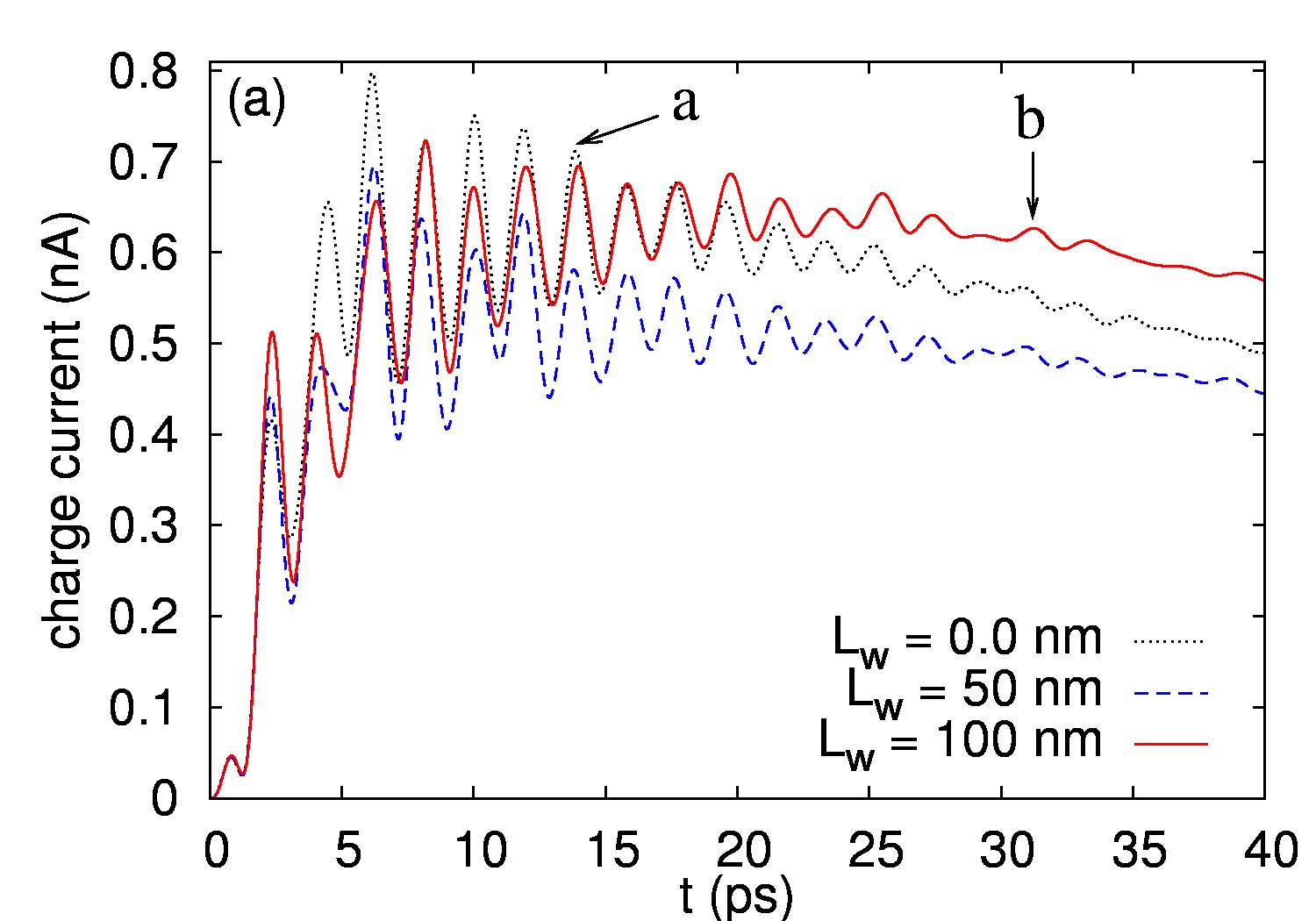}
 \includegraphics[width=0.45\textwidth]{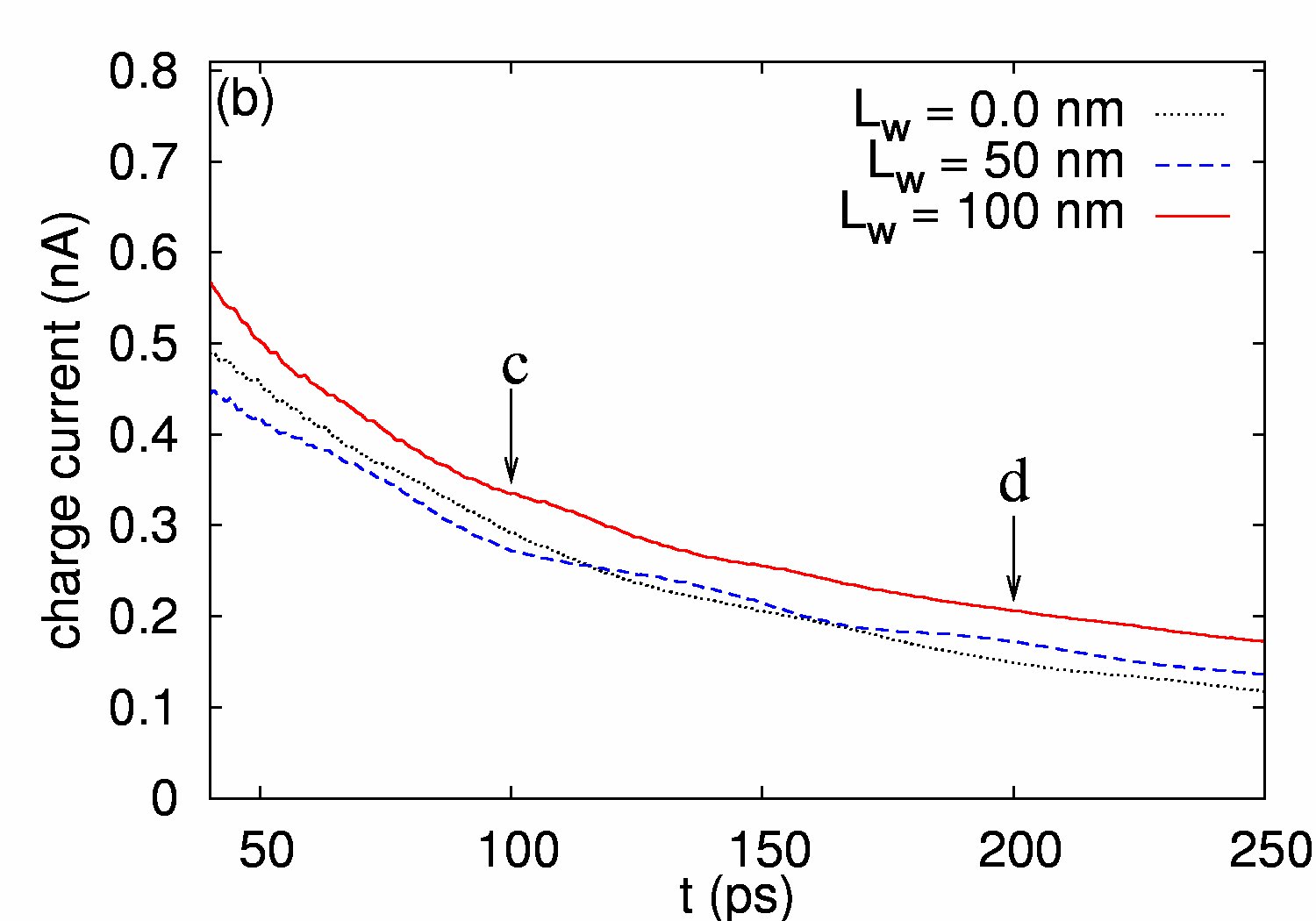}
\caption{(Color online) The interacting net current $I_{Q,I}$ versus
time for $L_{\rm w} = 0$ (dotted black), $50$ (dashed blue), and
$100$~nm (solid red): (a) short-time response; (b) long-time
response for $B = 0.5$~T. The other parameters are the same as
\fig{I-B05}.} \label{ILw}
\end{figure}

In \fig{ILw}, we show the interacting net charge current as a
function of time for the case of magnetic field $B=0.5$~T with
different size of coupling window $L_{\rm w} = 0$ (dotted black),
$50$ (dashed blue), and $100$~nm (solid red).   In the short-time
response regime, shown in \fig{ILw}(a), the quantum interference
dominates the time-dependent charge current feature with rapid
oscillation.  The oscillation amplitude and frequency of the
time-dependent charge current remain similar for the cases with
different window size, this similarity is because the quantum
interference oscillation behavior is mainly due to the multiple
scattering in the transport direction, and interference of
subbands in the semi-infinite leads.  In the transient
switching-on regime $t\leq 0.5$~ps, the charge current for both the
cases of short $L_{\rm w} = 50$~nm and long window $L_{\rm w} =
100$~nm are similar to the case without a window $L_{\rm w} =
0.0$~nm exhibiting the response time of the system from an isolated
system to an open system. Later on, the charge current for the case
of short window is suppressed by $0.4$~nA, while the charge current
is enhanced for the case of long window by $0.8$~nA. It should be
noted that this quantitative feature is different when the $e$-$e$
interaction effect is ignored, in which the charge current is almost
the same for the cases without window $L_{\rm w} = 0.0$~nm and long
window $L_{\rm w} = 100$~nm, however the charge current is
suppressed by $1$~nA for the case of short window $L_{\rm w} =
50$~nm (not shown).

In the long-time response regime, shown in \fig{ILw}(b), the
time-dependent charge current displays slow oscillations and
approaches to a steady current within $0.1-0.2$~nA. It is shown that
the steady current is enhanced for the case of long window $L_{\rm
w} = 100$~nm due to the Coulomb interaction.  However, the Coulomb
interaction for the case of short widow is not significant on the
time-dependent charge current in comparison with the pure finite
length DW system without window coupling.  When the Coulomb
interaction is ignored, the steady currents of the short and the
long window are both suppressed (not shown).  This demonstrates
again the dynamics of the time-dependent charge current in the
long-time response regime is significantly affected by the Coulomb
interaction.

\begin{figure}[tb]
 \includegraphics[width=0.15\textwidth]{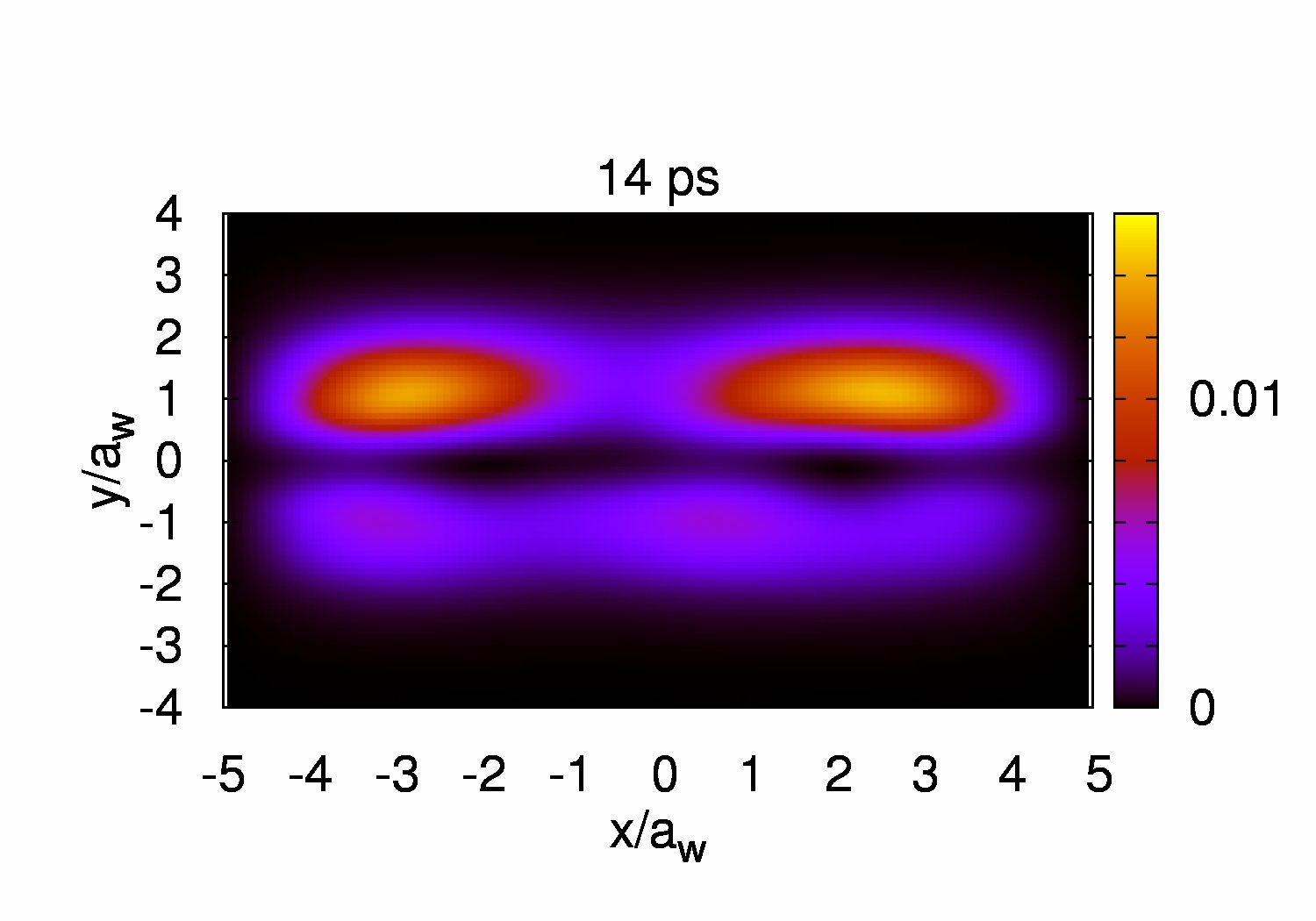}
 \includegraphics[width=0.15\textwidth]{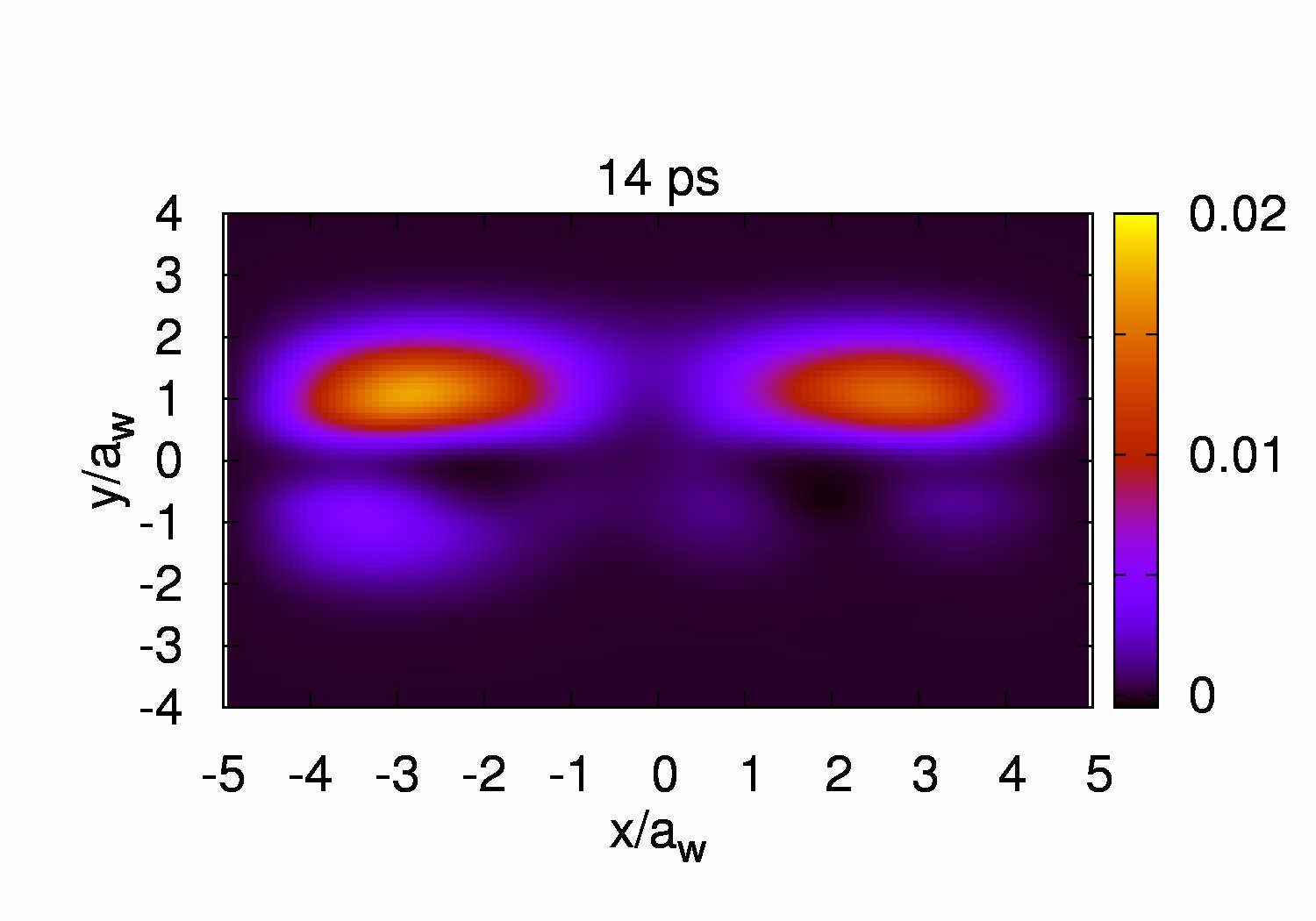}
 \includegraphics[width=0.15\textwidth]{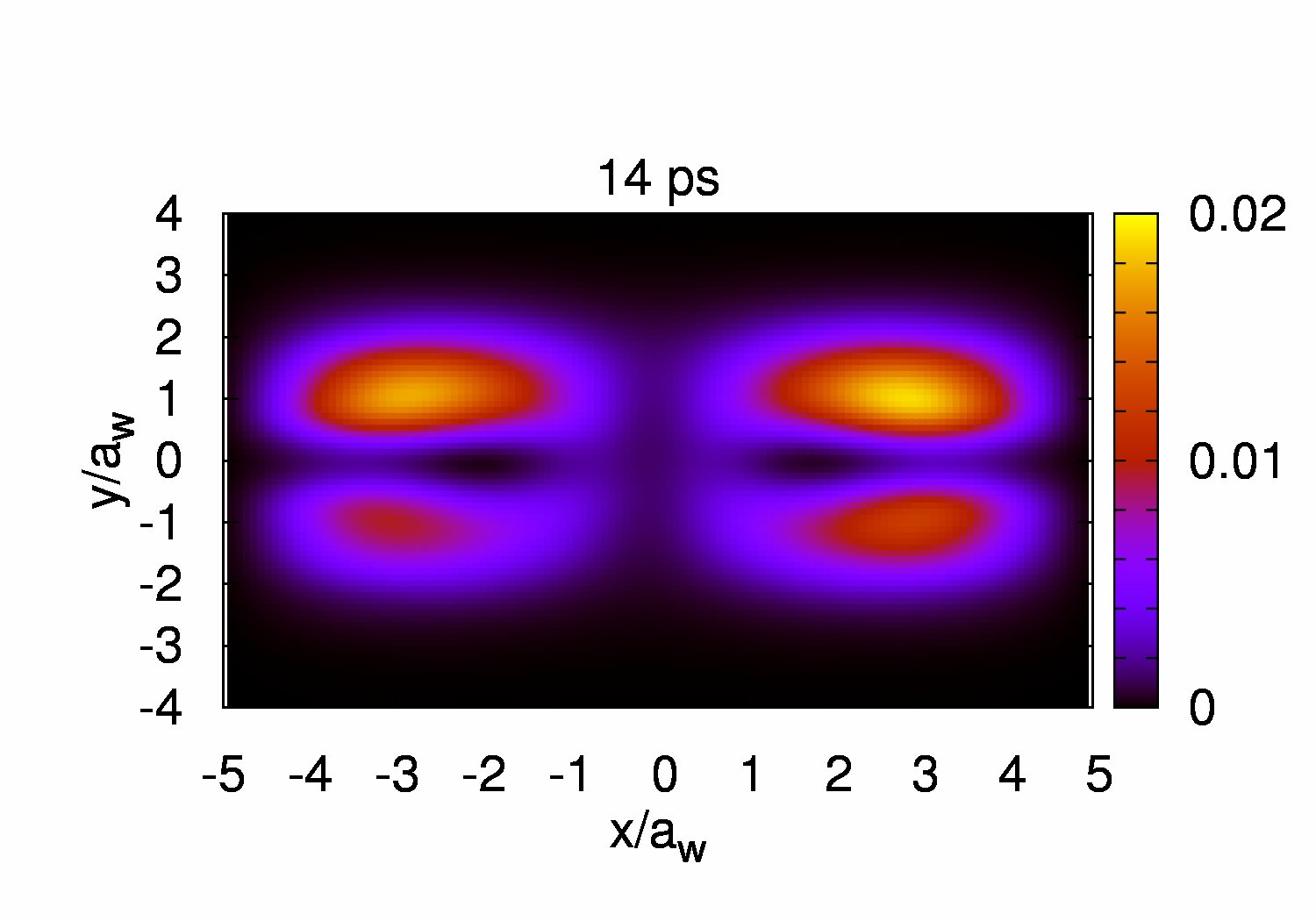} \\
 \includegraphics[width=0.15\textwidth]{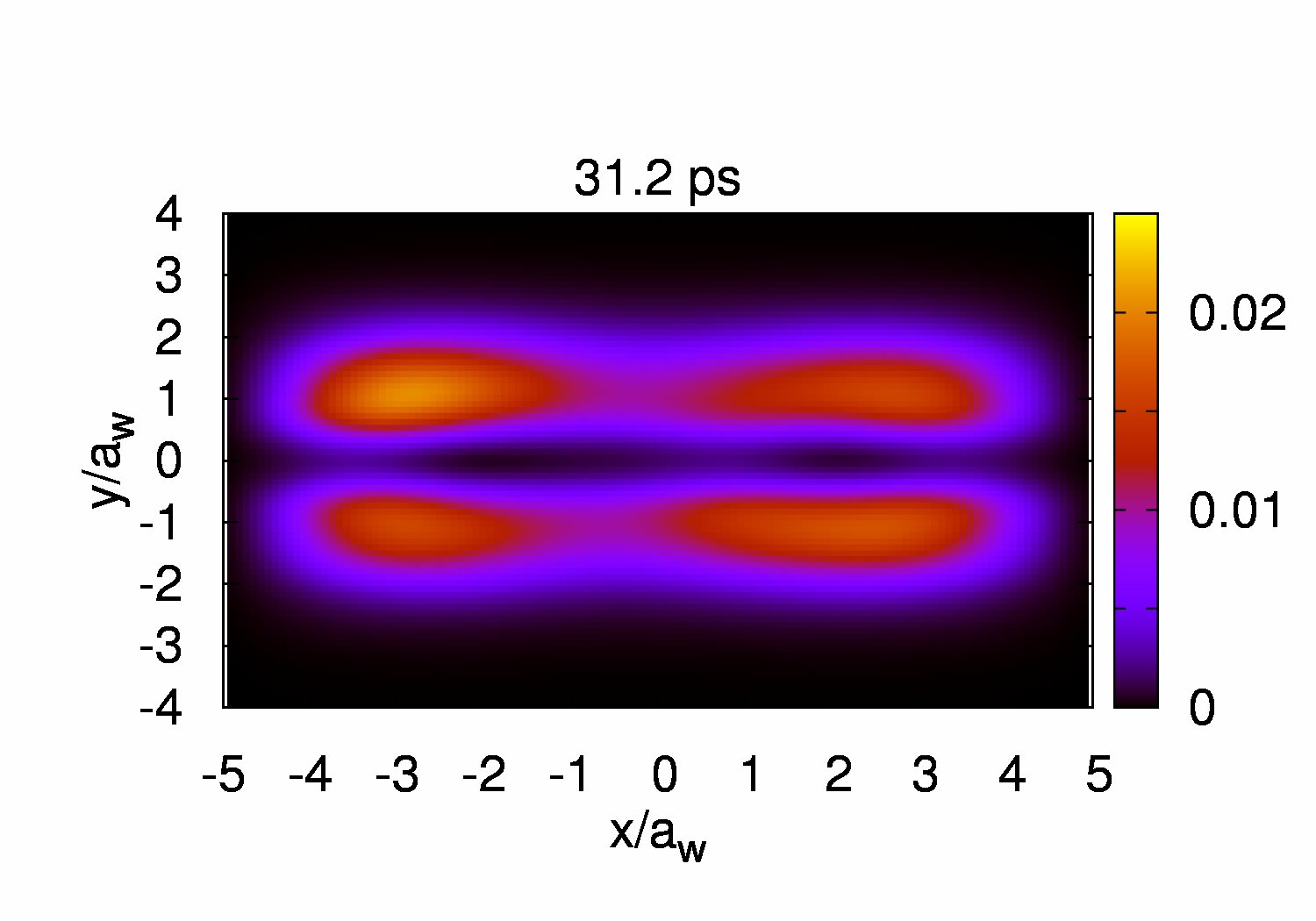}
 \includegraphics[width=0.15\textwidth]{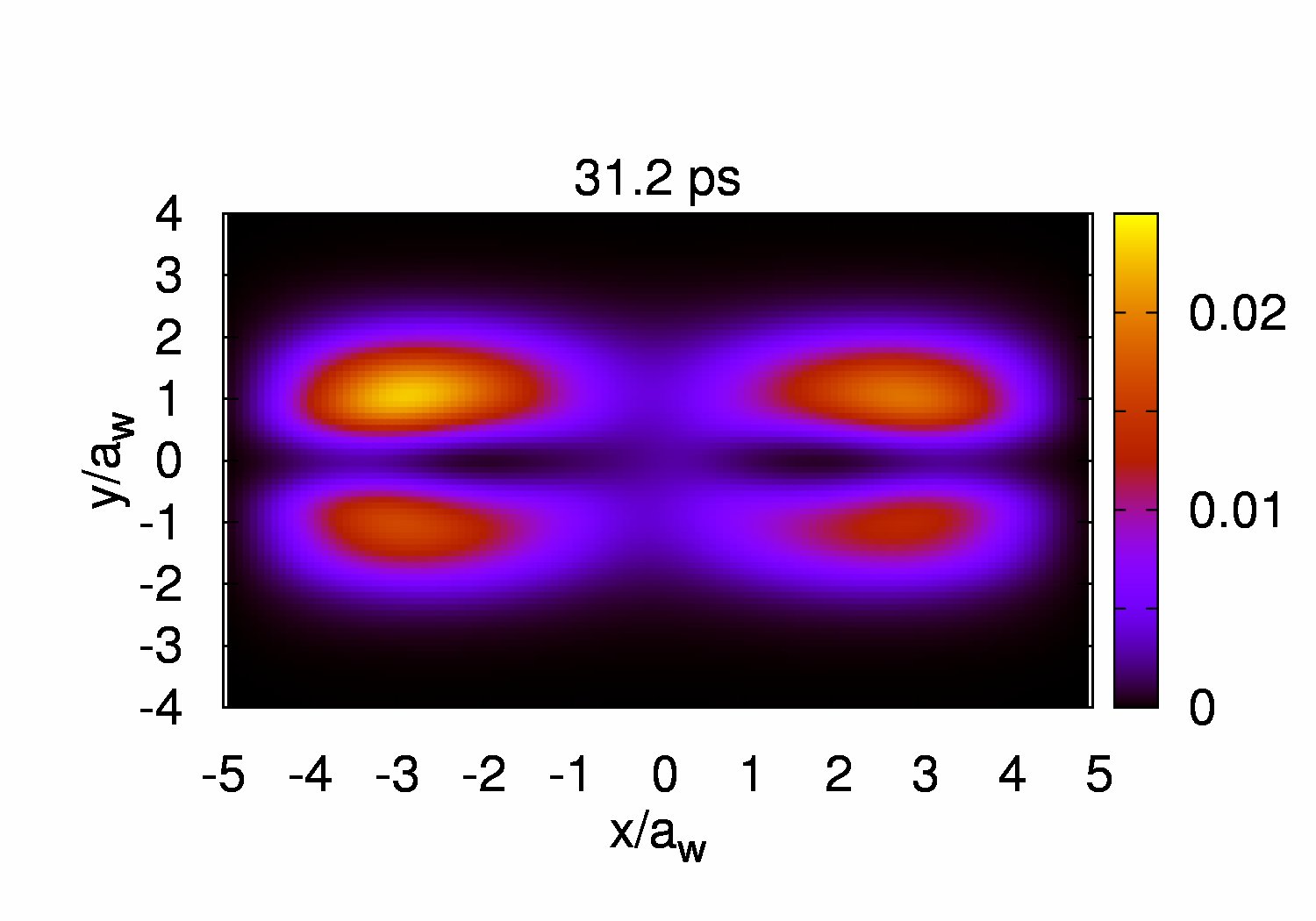}
 \includegraphics[width=0.15\textwidth]{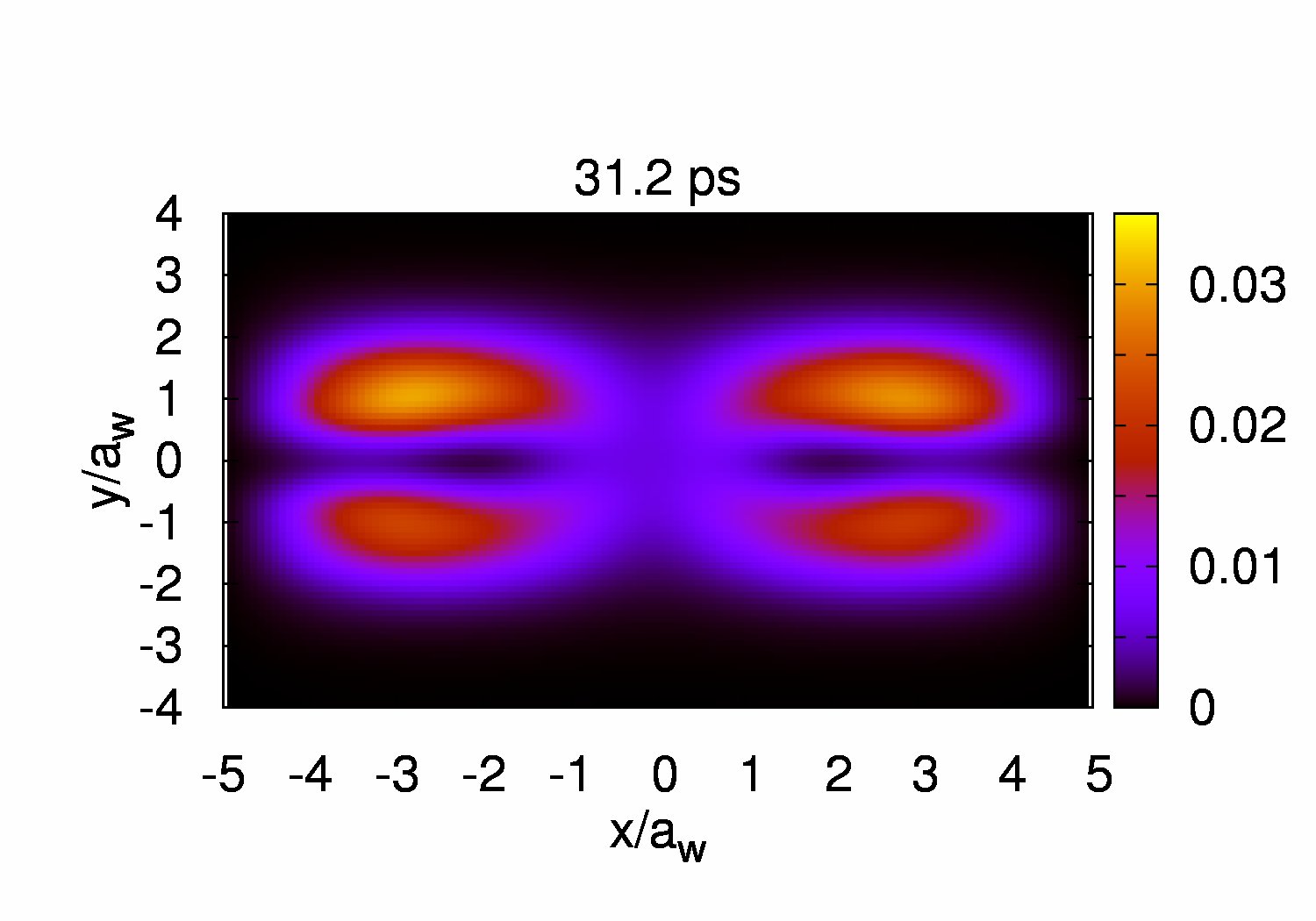} \\
 \includegraphics[width=0.15\textwidth]{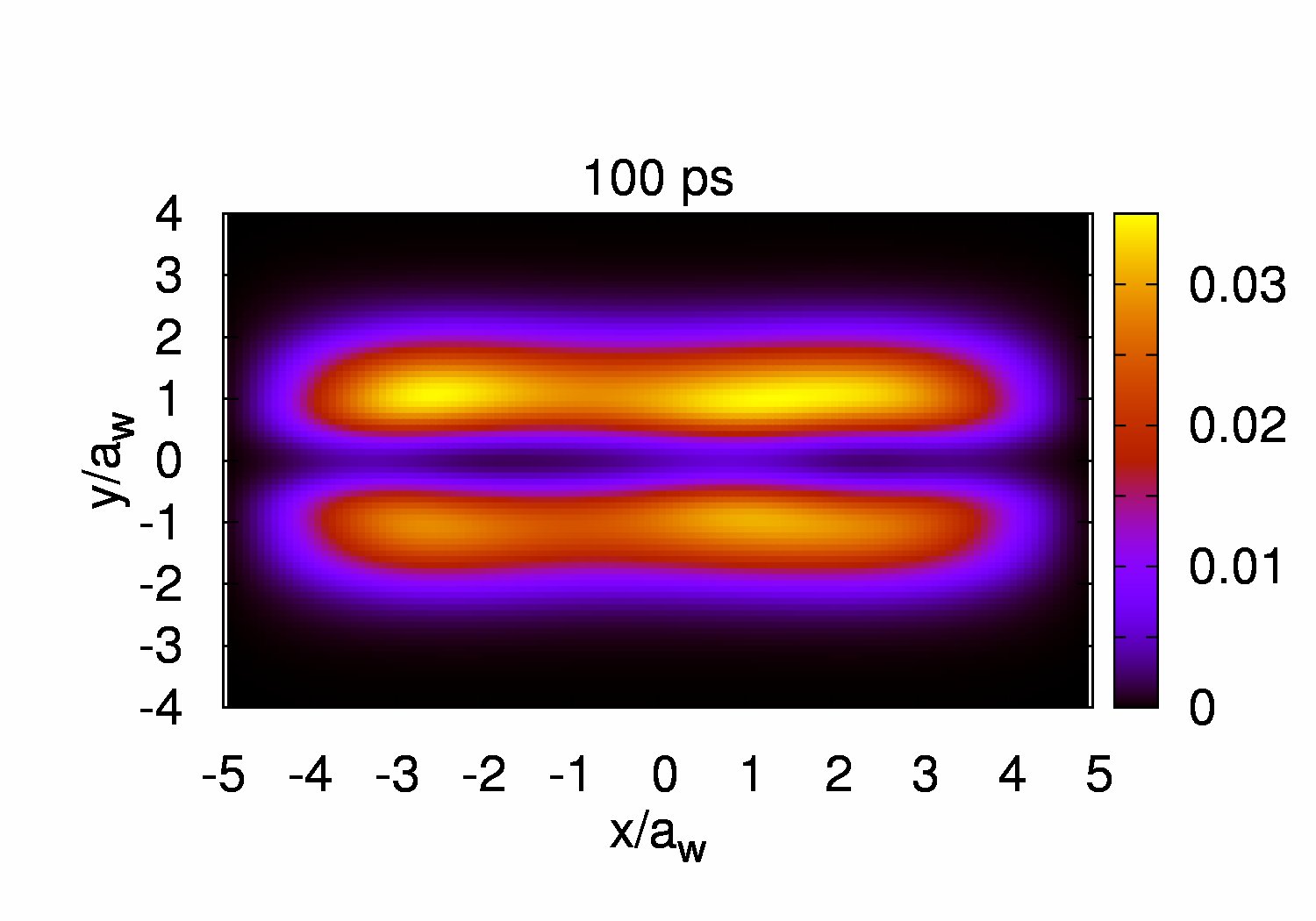}
 \includegraphics[width=0.15\textwidth]{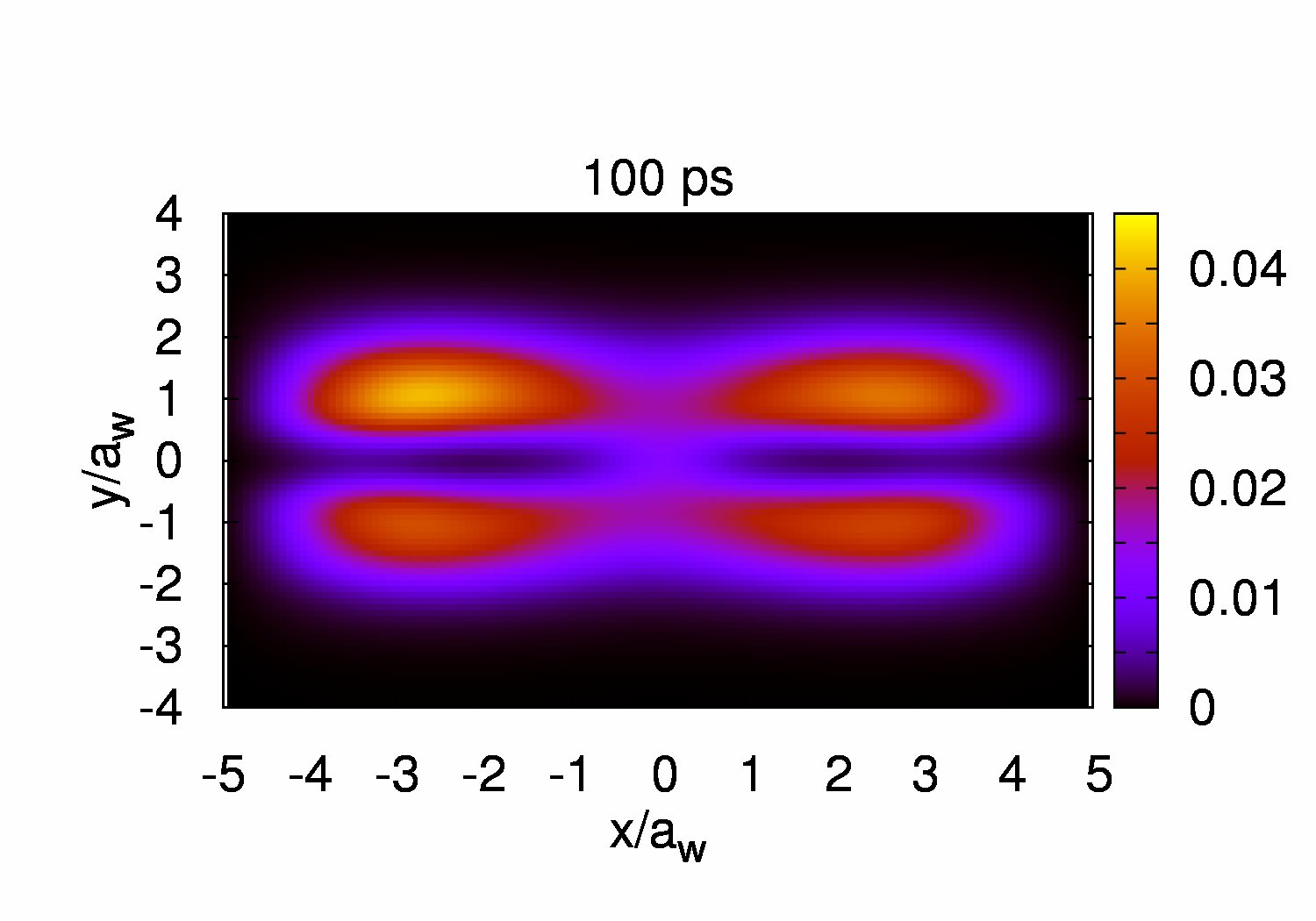}
 \includegraphics[width=0.15\textwidth]{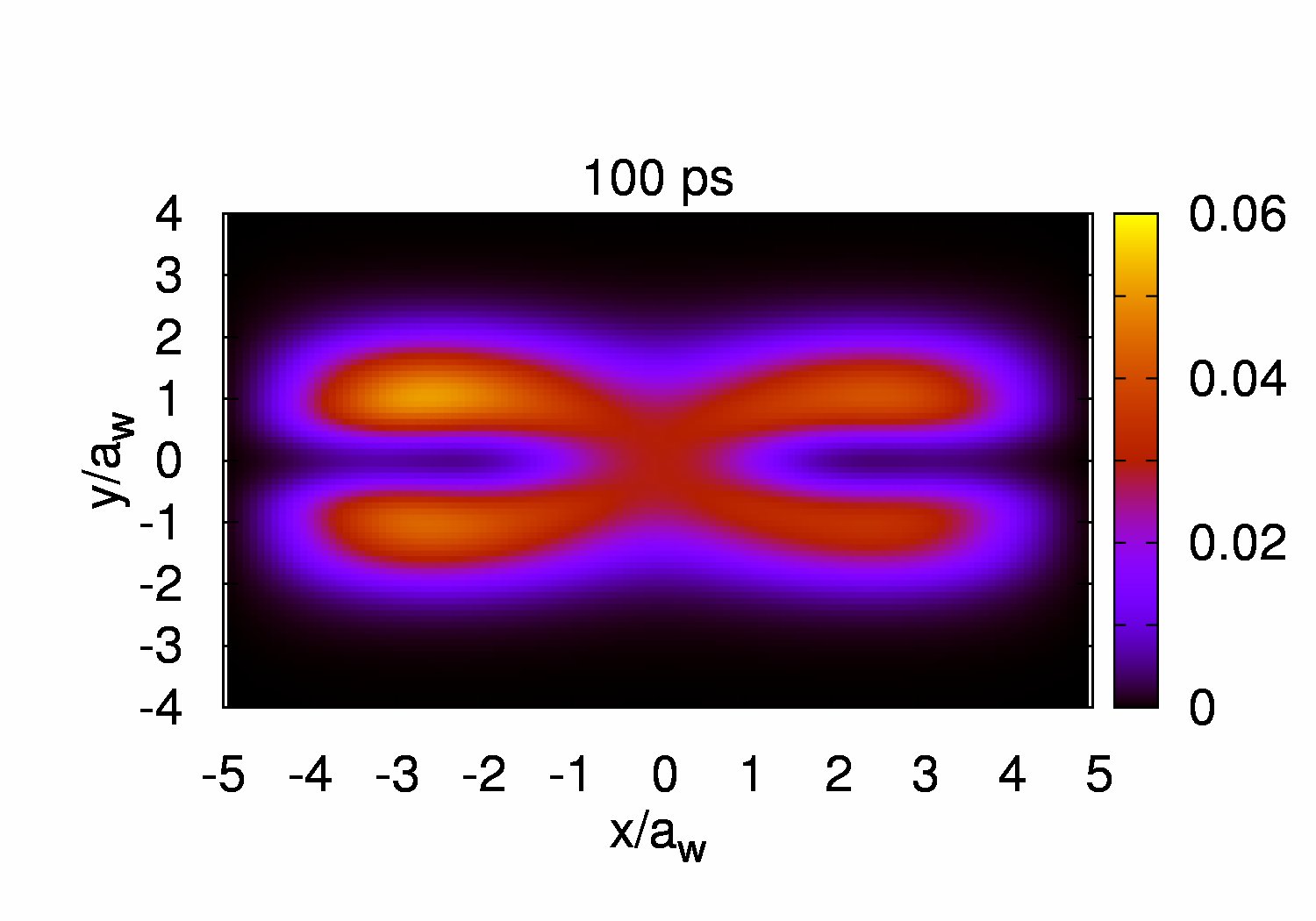} \\
 \includegraphics[width=0.15\textwidth]{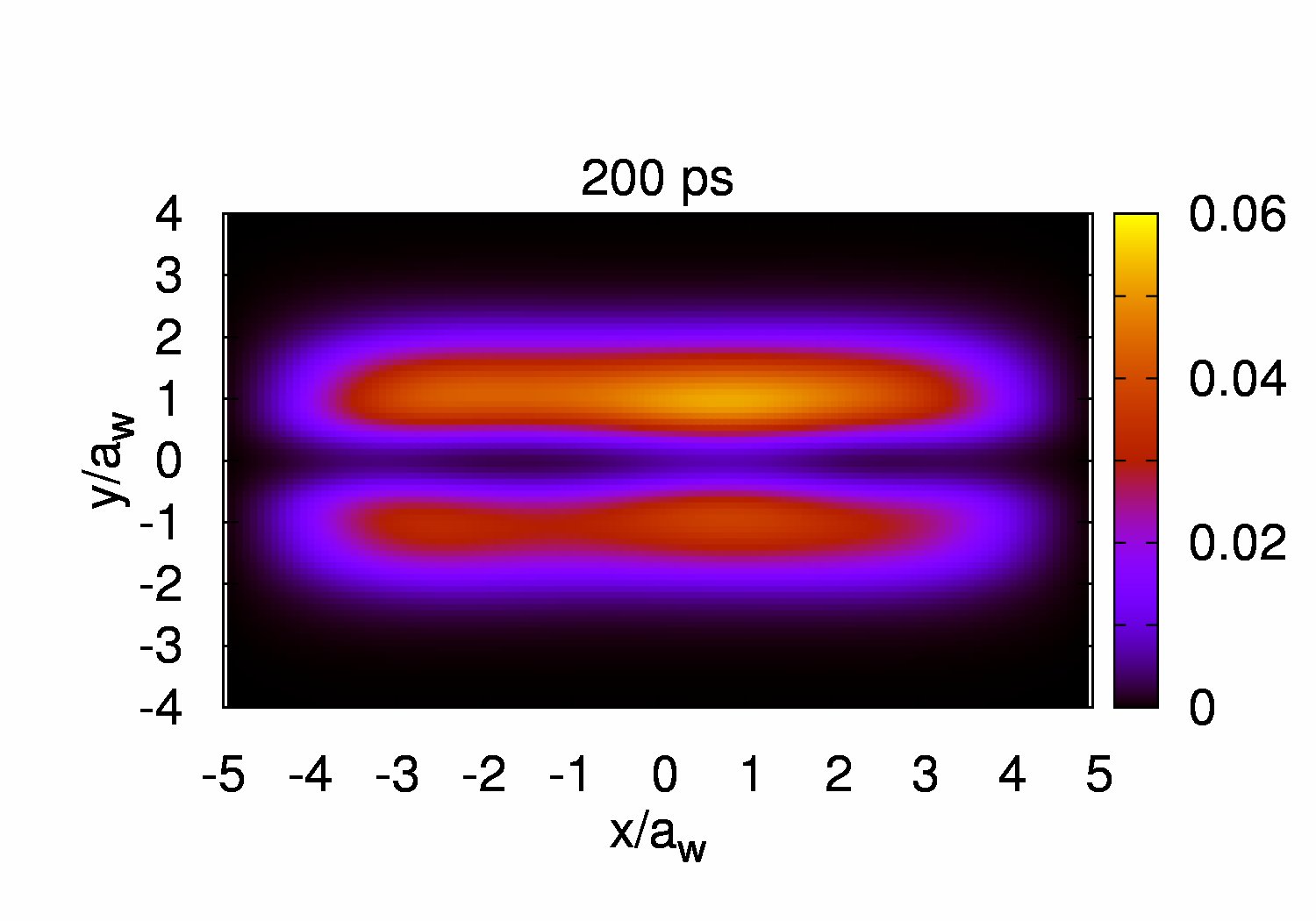}
 \includegraphics[width=0.15\textwidth]{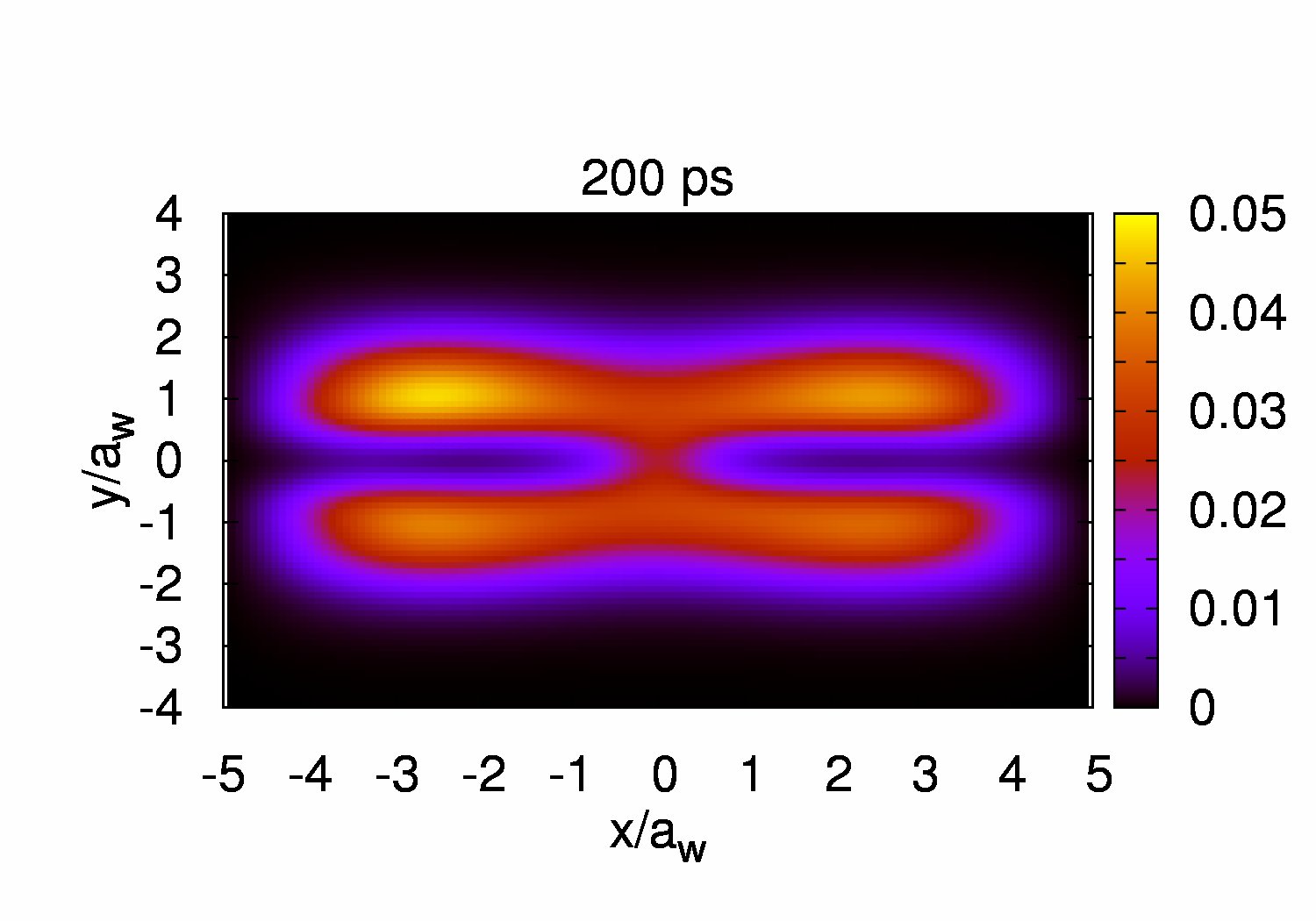}
 \includegraphics[width=0.15\textwidth]{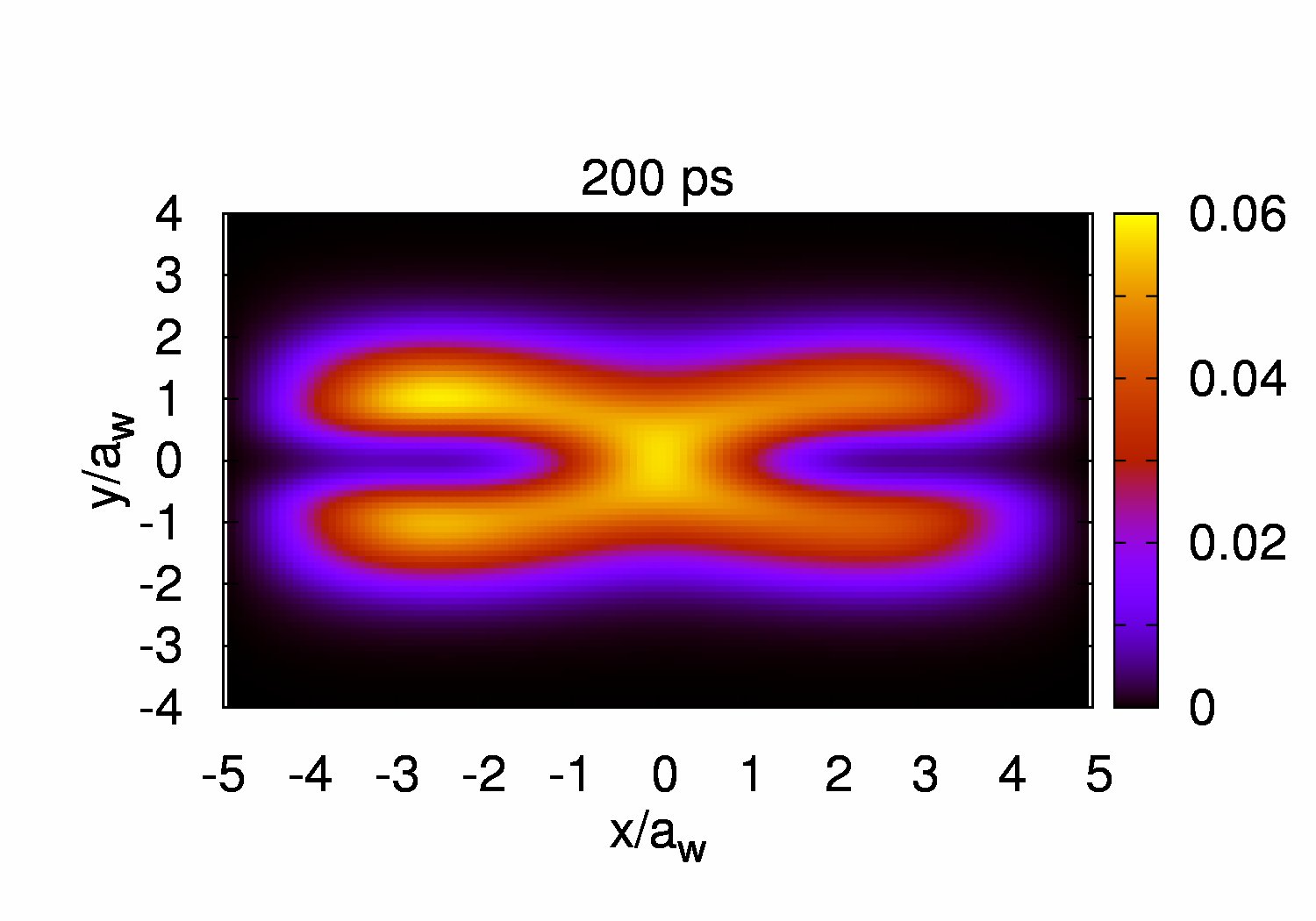} \\
\caption{(Color online) The spatial distribution of the interacting
many-electron charge density with different coupling window: $L_{\rm
w} = 0.0$~nm (left), $L_{\rm w} = 50$~nm (middle), and $L_{\rm w} =
100$~nm (right) at time $t=14$, $31.2$, $100$, and $200$~ps.
$B=0.5$~T, and the other parameters are the same as \fig{I-B05}.}
\label{QL}
\end{figure}

To investigate how the window size affects the transport dynamics,
in \fig{QL} we present the spatial distribution of the many-electron
charge at $t=14$, $31.2$, $100$, and $200$~ps, labeled by
$\bf{a}$-$\bf{d}$ in \fig{ILw}, respectively.  It is clearly seen
that, for both short and long window, the electrons
perform inter-wire backward scattering in the short-time response
regime (say, $t=14$ and $31.2$~ps), while the electrons are allowed
to perform inter-wire forward scattering in the long-time response
regime (say, $t=100$ and $200$~ps).  This means that the former
quantum interference dominant short-time response regime, the
electrons favor the inter-wire backward scattering; while the latter
Coulomb interaction dominant long-time response regime, the
electrons favor inter-wire forward scattering.  The many-electron
charge density is monotonically increased in time. Furthermore, it
is demonstrated that increasing window size can enhance not only the
inter-wire scattering feature, but also the local charge
accumulation at the coupling window.

\section{Concluding Remarks}\label{Sec:IV}

To conclude, we have performed a numerical
calculation of the time-dependent electric current and spatial
charge distribution through a window-coupled parallel double quantum
wire system based on GQME formalism including the electron-electron
Coulomb interaction with the ``exact diagonalization'' method, and
without resorting to the commonly used Markovian approximation.
We have analyzed transient
currents and their dependence on various parameters of the system
with a certain initial configuration and time-dependent switching-on
coupling to the leads. For a given coupling window, we have
demonstrated time-dependent transport properties of the
noninteracting and the interacting DW systems. Applying an
appropriate magnetic field, we have found a short-time response regime
dominated by quantum interference and inter-wire
backward scattering. Moreover, the Coulomb
correlation is significantly enhanced in the long-time response
regime,\cite{Vidar2010} and the inter-wire forward scattering through the coupling
window dominates the dynamical transport properties.  The conceived
mesoscale window-coupled DW system could serve as an elementary
quantum device for sensitive spectroscopy tools for electrons and
quantum information processing by controlling the coupling window
and the applied magnetic field.


%
\begin{acknowledgments}
This work was supported by the Research and Instruments Funds
Icelandic; the Research Fund of the University of Iceland; the
Icelandic Science and Technology Research Programme for Post-genomic
Biomedicine, Nanoscience and Nanotechnology; and the National
Science Council in Taiwan through Contract No.
NSC97-2112-M-239-003-MY3.
\end{acknowledgments}

%
%

%
%
\end{document}